\documentclass{article}

\usepackage{arxiv}

\usepackage[utf8]{inputenc}
\usepackage[T1]{fontenc}
\usepackage{hyperref}
\usepackage{url}
\usepackage{booktabs}
\usepackage{amsfonts}   
\usepackage{microtype}
\usepackage{graphicx}
\graphicspath{ {./figures/final/} }

\usepackage{makecell}
\usepackage{tabularx}
\usepackage{ragged2e}
\usepackage{adjustbox}
\usepackage[english]{babel}
\usepackage{tikz}
\usepackage[table,xcdraw]{xcolor}
\usepackage{lscape}
\colorlet{jaws_comments}{.}
\usepackage[textsize=footnotesize, colorinlistoftodos]{todonotes}

\title{Rethinking Code Review in the Age of AI: A Vision for Agentic Code Review}

\renewcommand{\headeright}{}
\renewcommand{\undertitle}{}
\author{
  H\"{u}seyin \"{O}zg\"{u}r Kamal{\i} \\{}
  Department of Software Engineering \\{}
  Ankara University \\{}
  Ankara, Turkey \\
  \texttt{22290405@ogrenci.ankara.edu.tr} \\
  \href{https://orcid.org/0009-0009-9864-9513}{ORCID: 0009-0009-9864-9513}
  \And
  Erdem Tuna \\{}
  Microsoft \\{}
  Ankara, Turkey \\
  \texttt{erdemtuna@microsoft.com} \\
  \href{https://orcid.org/0000-0001-7137-6361}{ORCID: 0000-0001-7137-6361}
  \And
  Vahid Haratian \\{}
  Department of Computer Engineering \\{}
  Bilkent University \\{}
  Ankara, Turkey \\
  \texttt{vahid.haratian@bilkent.edu.tr} \\
  \href{https://orcid.org/0009-0001-3048-9586}{ORCID: 0009-0001-3048-9586}
  \And
  Eray T\"{u}z\"{u}n \\{}
  Department of Computer Engineering \\{}
  Bilkent University \\{}
  Ankara, Turkey \\
  \texttt{eraytuzun@cs.bilkent.edu.tr} \\
  \href{https://orcid.org/0000-0002-5550-7816}{ORCID: 0000-0002-5550-7816}
}

\begin{document}
\maketitle

\begin{abstract}

Code review has evolved for decades, from informal peer checking to today's pull request (PR) workflows, yet it remains a largely manual and cognitively demanding process.
The rise of Artificial Intelligence (AI) coding assistants has intensified this challenge: while these tools increase code production velocity, they also expand the volume of code requiring review, turning code review into a growing bottleneck.
Current AI support in code review remains fragmented, with tools focusing on isolated tasks such as reviewer recommendation, PR description generation, or comment suggestion rather than the end-to-end PR review workflow.
We address this gap by treating review effectiveness as an outcome of the full code review lifecycle rather than a single stage, proposing a framework that carries context across stage boundaries.
We propose a future vision for code review in which reviewers transition from manual inspectors into supervisory operators of agents.
In this vision, staged, AI-powered workflows aim to align the pace of code generation with shared understanding and accountable engineering.
In this paper, we review the historical evolution of code review practices, identify challenges in traditional code review systems, and examine the shift driven by large language models (LLMs) and agentic AI systems.
We then present a vision for an AI-powered code review workflow combining specialized agents with human-controlled quality gates.
Our framework spans five stages: \textit{PR Creation}, \textit{PR Augmentation}, \textit{Reviewer Selection}, \textit{AI-Assisted Code Review}, and \textit{PR Retrospective}, with humans retained at key decision points to preserve judgment, accountability, and team-level understanding.
We then identify open challenges for responsible adoption, including reliability, bias, privacy, automation bias, transparency, and evaluation.
We offer a research agenda targeting evaluation benchmarks, bias auditing, and governance models for responsible human-AI collaboration in software engineering.
\end{abstract}


\newpage
\section{Introduction}
\label{introduction}


Code review is a common practice in modern software development. Developers examine source code produced by peers to identify defects and facilitate knowledge transfer~\cite{Bacchelli2013}. Beyond defect detection, code review also supports enforcing coding standards, mentoring junior contributors, and disseminating architectural knowledge across teams~\cite{Sadowski2018}. These objectives have converged across open-source and industrial practice into a shared set of expectations for what review should accomplish~\cite{Rigby2013}. In contemporary workflows, this activity is operationalized through pull requests (PRs) on platforms such as GitHub~\cite{github2025}, GitLab~\cite{gitlabFinallyEntire}, and Bitbucket~\cite{bitbucketBitbucketSolution}, which integrate version control, discussion, and continuous integration into a unified review environment~\cite{Gousios2014}. Nearly every change of consequence now passes through this review pipeline. Yet as Artificial Intelligence (AI) assisted development reshapes how code is produced, the core structure of the review pipeline through which that code must pass has remained largely unchanged.

Despite its central role, PR-based review faces persistent challenges. PR reviewers often lack the rationale and behavioral context needed to evaluate a change, and understanding this context remains PR reviewers' primary challenge~\cite{MacLeod2018}. Reviewer workload~\cite{Baysal2016}, change complexity~\cite{Ram2018, Kononenko2016}, and code churn~\cite{nagappan2005use} shape review quality in ways that disconnected tools do not account for. Matching the right reviewer to the right change is another recurring source of review friction. These challenges are reinforced by recurring malpractices such as incomplete PR descriptions, poor reviewer assignment, and unconstructive feedback, which accumulate technical debt, erode code quality and reduce the effectiveness of code review~\cite{Doan2022, McIntosh2016}. These challenges are not new. What changes now is the pace and provenance of the code entering review.

AI does not only help produce code. It also rebounds on the review process through which that code must pass. AI coding assistants accelerate individual coding tasks by more than 50\%~\cite{Peng2023}, but this gain does not propagate uniformly through the workflow. Coordination time for integration grows faster than individual output~\cite{Song2024}, and AI-generated contributions themselves require more review iterations than human-written ones~\cite{Zhong2026}. When AI assists review, PR reviewers surface more low-severity issues but not additional high-severity defects~\cite{Tufano2025}, which suggests that automated support can pull attention toward the easier problems. Developers can also over-rely on AI output~\cite{Sabouri2025, Khati2025}, and the pace of AI adoption can outrun the development of review and debugging skills~\cite{Fawzy2025}. AI reduces the cost of writing code. At the same time, it raises the cost and the stakes of reviewing that code~\cite{Abraho2025}. Under these conditions, code review is no longer only a productivity bottleneck. It is the primary control surface for the quality and accountability of AI-produced code.

Prior work~\cite{Davila2021, Badampudi2023} has advanced individual stages of the review process. Approaches address review comment generation~\cite{Tufano2022, Lu2023, hong2024commentfinder}, PR description generation~\cite{Xiao2024}, reviewer recommendation~\cite{Wang2024}, and code comprehension during review~\cite{Nam2024}. Recent multi-agent systems coordinate activity within the review phase itself~\cite{Ren2025, Tang2024}. Each of these contributions improves one stage of the workflow. These stage-level advances, however, do not compose on their own. A helpful review comment still depends on a PR whose rationale was written down~\cite{MacLeod2018}. Reviewer matching relies on behavioral context reaching the reviewer~\cite{MacLeod2018}. Future changes depend on lessons from prior reviews being written down. These dependencies cross the tool boundaries between stages, and stage-level work alone cannot resolve them.

We argue that review effectiveness must be treated as an outcome of the full code review process lifecycle rather than as the result of a single review stage. This reframing changes what a code review system is required to do. Such a system carries context across stage boundaries, so that the output of one stage becomes usable input to the next. It places AI where it reduces coordination cost rather than where it absorbs decisions that belong to human reviewers. And it keeps the lifecycle legible over time, so that evidence from past reviews can inform later ones. We develop this into a concrete framework of coordinated stages in Section~\ref{sec:ai-powered-framework}.

This vision paper makes three conceptual contributions. First, we argue that AI-accelerated code production amplifies rather than mitigates existing shortcomings of PR-based review. Second, we argue that improvements scoped to single stages of the review process cannot, on their own, produce effective reviews across the full code review process lifecycle. Third, we propose an AI-powered code review framework that enables this reframing through five coordinated stages. We also outline a research agenda for evaluating review systems, including the metrics, study designs, and questions of human and AI authority such evaluation will require.

The rest of this paper is organized as follows. Section~\ref{sec:evolution-of-code-review-practices} reviews the history of code review from its origins to contemporary practices. Section~\ref{sec:traditional-code-review-systems} analyzes traditional code review systems and their associated challenges. Section~\ref{sec:ai-powered-framework} presents our proposed AI-powered code review framework. Section~\ref{sec:discussion} discusses potential challenges, risks, and limitations alongside implications for practitioners and researchers. Finally, Section~\ref{sec:conclusion} concludes the paper.

\section{Evolution of Code Review Practices}
\label{sec:evolution-of-code-review-practices}

This section examines the behavioral evolution of code review through five distinct eras.
Each era is defined by how practitioners conducted code review, including the methodologies, processes, and social conventions that characterized the period.
This practice-based framework emphasizes shifts in review philosophy, from informal problem-solving to structured inspection, from heavyweight meetings to lightweight asynchronous exchange, and from purely human evaluation to automation-assisted workflows.
Table~\ref{tab:code-review-evolution} summarizes these practice dimensions across all five eras.

\subsection{Ad Hoc Code Review Era}
\label{subsec:ad-hoc-code-review-era}

During the earliest period of computing (1940s-1960s), systematic code review practices were largely absent.
Software engineering had not yet emerged as a formal discipline, and standardized quality assurance (QA) approaches did not exist~\cite{Wirth2008}.
Programming during this era was characterized by highly individualized, often solitary work.
The code was submitted manually using punched cards or paper listings, and the success of a program typically depended on individual expertise rather than collaborative verification~\cite{Backus1980}.

Although small programming teams existed, collaboration typically involved specialized individual contributions rather than a collective examination of the source code.
When informal consultation occurred, it was reactive and unstructured: programmers might seek help from available colleagues when encountering difficulties, with feedback conveyed verbally or through handwritten marginal notes on printouts~\cite{Wirth2008}.
There were no formalized roles, structured processes, or approval gates.

As project complexity increased throughout the 1950s and 1960s, the absence of QA practices contributed to mounting difficulties in maintaining software quality, preventing defects, and ensuring maintainability.
The 1968 NATO Software Engineering Conference~\cite{Naur1969} formally articulated the \textit{software crisis}, recognizing that undisciplined development methods were inadequate for increasingly complex systems.
During this period, Weinberg~\cite{Weinberg1971} proposed the concept of \textit{egoless programming}, advocating that programmers conduct peer reviews \textit{in a friendly and collegial way} without hierarchical barriers. This was a prescription for how teams \textit{should} work, rather than a description of the dominant practice.
This recognition of the need for collaborative QA established the intellectual foundation for formalized inspection methodologies, setting the stage for Fagan's systematic approach in the following decade.

\subsection{Formal Inspection Era}
\label{subsubsec:formal-inspection-era}

The Formal Inspection Era (1970s-1990s) marked the first rigorous, methodological approach to code review, establishing practices that would influence all subsequent review methodologies.
In 1976, Fagan~\cite{Fagan1976} introduced a structured, role-based approach. 
In this process, the \emph{designer} is the person designing the program flow, the code artifacts are created by the \textit{coder}, the \textit{moderator} leads the inspection process as the coach, and the \textit{tester} writes and tests the resulting product.
The five-step process (overview, preparation, inspection meeting, rework, and follow-up) provided a repeatable framework with measurable outcomes.

Fagan~\cite{Fagan1976} reported that inspections could detect the majority of errors before testing began, in some cases up to 80\%, an improvement over ad-hoc review approaches.
The results also suggested notable gains in development efficiency, though the improvements varied across contexts and have been subject to investigation.
Other studies presented similar findings in practical and industrial settings~\cite{Ackerman1989, Russell1991}. 
Exploring further, Fagan's later work~\cite{Fagan1986} documented lessons learned from applying inspections across IBM projects over the years. 
First, a \emph{planning} step was prepended to the development process in~\cite{Fagan1976}, officially establishing it as a six-step development process consisting of planning, overview, preparation, inspection, rework, and follow-up.
Second, the work distinguished formal inspections from the more \emph{relaxed} review style known as a \emph{walkthrough}, which was found to be less efficient and lacked repeatable data collection.
Additionally, the paper addressed practical concerns such as determining when code was ready for inspection and when it could safely proceed to the next development phase by implementing objective entry and exit criteria.
The practice dimensions of this era emphasized thoroughness over speed. 
Trained moderators coordinated small, specialized teams conducting synchronous face-to-face meetings.
Written checklists were utilized during the inspection, and defects were recorded.
Furthermore, no artifact was allowed to pass until all rework was verified against the checklists.

By the early 1990s, the rigidity of Fagan's process prompted researchers to explore alternatives, that would eventually pave the way for lightweight review.
Parnas and Weiss~\cite{Parnas1985} focused on more opinionated reviewer selection and increased activity in the design review process, before writing code.
Knight and Myers~\cite{Knight1991} introduced phased inspections to improve review quality by decomposing the process into manageable phases.
Each phase targeted a specific quality property by leveraging a software-based tool.
Similarly, Brothers and Sembugamoorthy~\cite{Brothers1990,Sembugamoorthy1990} introduced a collaborative software-based system where reviewers could submit comments, propose changes, and track revisions. By centralizing reviewer assignments and feedback, it shifted inspections from manual activities to integrated, tool-supported practices.

\subsection{Lightweight Peer Review Era}
\label{subsec:lightweight-peer-review-era}
The Lightweight Peer Review Era (1990s-2000s) fundamentally reshaped how code review was conducted, trading the rigor of formal inspections for velocity and accessibility.
Starting in the early 1990s, and building on the critiques of formal inspection rigidity from the previous era, Votta~\cite{Votta1993} directly questioned whether inspection meetings were necessary at all.
His empirical study found that the majority of defects were identified during individual preparation rather than during synchronous meetings, suggesting that the coordination overhead of formal gatherings provided limited additional value.
Johnson and Tjahjono~\cite{Johnson1998} extended this investigation, comparing meeting-based and non-meeting-based inspection methods and finding no significant difference in defect detection rates.
Together, these findings challenged a core assumption of Fagan's methodology and opened the door to asynchronous, tool-mediated review.

Open-source communities provided practical validation of this shift.
Projects such as the Linux kernel and Apache required review mechanisms for geographically distributed contributors who could not attend synchronous meetings~\cite{Raymond2001, Mockus2002, Herbsleb2000}.
Concurrently, the Agile movement~\cite{AgileAlliance2001} emphasized rapid iteration and minimal ceremony, making formal inspections seem incompatible with iterative development cadences.
Together, these forces demonstrated that effective review could occur without the heavyweight process.

In practice, review during this era followed the rhythms of mailing list communication.
A developer would prepare a patch, a text file representing changes to the codebase, and post it to the project's mailing list~\cite{Raymond2001}.
Any interested contributor could then examine the diff and respond.
Unlike formal inspections with assigned roles, reviewers were often self-selected volunteers who brought relevant expertise or simply had time to contribute.
Discussion unfolded asynchronously through email threads, with reviewers quoting specific code segments and suggesting alternatives, the author would revise and resubmit until the maintainer judged the contribution ready for integration~\cite{Mockus2002}.
This trust-based model replaced formal approval gates with community consensus, relying on the maintainer's judgment rather than structured verification.
Version control systems (VCSs) such as CVS~\cite{Berliner1990} provided the infrastructure for generating and applying patches, but no specialized review platforms existed.

Empirical studies of open-source peer review documented similar patterns emerging across different projects.
Mockus et al.~\cite{Mockus2002} examined Apache and Mozilla development, finding that distributed contributors coordinated effectively through email-based review and maintainer-driven integration.
Rigby et al.~\cite{Rigby2008} extended this analysis for Apache, demonstrating that reviews were small, occurred frequently during development, were asynchronous rather than meeting-based, and emphasized knowledge transfer alongside defect detection.
This convergence suggested that the lightweight model represented a natural evolution of effective practices across contexts.
However, email-based review had its own limitations, including information overload and difficulty discovering relevant discussions~\cite{Rigby2011Paper}, challenges in tracking review progress~\cite{Rigby2011PhD}, and a lack of structured audit trails~\cite{Rigby2011PhD}.
These shortcomings motivated the development of web-based tooling that would characterize the next era.

\subsection{Integrated Code Review Era}
\label{subsec:integrated-code-review-era}
The Integrated Code Review Era (2000s-2010s) marked the widespread adoption of asynchronous, tool-supported peer review processes.
Google introduced an internal tool called Mondrian in 2006~\cite{Niall2006}, enabling reviews prior to committing to the master branch within a centralized repository. 
In the years that followed, a variety of tools emerged to support similar workflows. 
GitHub's launch in 2008 popularized the PR based software development model, streamlining code submission and review~\cite{Gousios2014}.
Other tools, both public and internal, such as Gerrit~\cite{GerritTeam2009}, ReviewClipse~\cite{Bernhart2010}, Phabricator~\cite{PhabricatorTeam2011}, and CodeFlow~\cite{CodeFlowTeam2012}, offered comparable capabilities, including inline comments, reviewer assignment, and merge gating.
During this era, the software engineering community established a workflow in which every code change underwent peer review, with early forms of automation integrated to support the process. 
Code review became a scalable collaboration mechanism, and automation improved efficiency by catching basic issues early. 
However, automation played a supporting role—project owners retained final decision-making authority, and many aspects of review remained inherently human.

In practice, reviews were typically asynchronous and conducted before merging. 
Developers submitted changes via PRs (or equivalent mechanisms), which reviewers accessed through web interfaces (e.g., Gerrit~\cite{GerritTeam2009}, Phabricator~\cite{PhabricatorTeam2011}), IDE plugins (e.g., ReviewClipse for Eclipse~\cite{Bernhart2010}), or desktop clients (e.g., CodeFlow~\cite{CodeFlowTeam2012}), depending on the tool.
Reviewer selection was often a manual and time-consuming task~\cite{Jeong2009}. 
Reviews were generally completed within hours to a day~\cite{Baysal2013}, and explicit approval was required before merging into the main codebase.
Notably, automated checks—such as build verification and basic test execution—began to integrate into review platforms. 
For example, Google integrated FindBugs into its review process, leading to the resolution of over 1,000 issues flagged by the tool~\cite{Ayewah2008}. 
This marked one of the earliest examples of static analysis tools supporting code review.

A defining characteristic of this era was the heavy use of branching in development workflows. 
Distributed VCSs (e.g., Git) enabled developers to work on isolated feature branches and submit PRs for integration. 
While this facilitated parallel development, it also introduced integration complexity. 
Bird and Zimmermann~\cite{Bird2012} examined the costs of extensive branching at Microsoft, finding that long-lived branches, termed "branchmania", delayed integration, with changes taking nearly nine additional days on average to reach the main codebase.
These "big bang'' merges were often error-prone and time-consuming. 
In response, many projects adopted best practices such as frequent forward-merging to keep branches in sync, batching PR merges, or maintaining hierarchical branch structures (e.g., maintenance, development, feature) to stage integrations in a controlled manner~\cite{Shihab2012}.

As PR-based development became the norm, project owners emerged as key quality gatekeepers.
Gousios et al.~\cite{Gousios2015} found that 75\% of GitHub projects mandated peer review for all contributions, with project owners prioritizing code quality, test completeness, and alignment with project goals. 
Popular projects often faced review overload, with dozens of open PRs requiring triage. 
To manage this, teams increasingly adopted CI gating—requiring builds and tests to pass before human review.
Vasilescu et al.~\cite{Vasilescu2014} found that over 90\% of analyzed GitHub projects had configured CI services, with automated checks reporting build status directly on PRs. 
These checks reduced integrator workload by catching build failures and trivial issues early.
Empirical evidence confirmed that this build-time signal surfaced integration failures on PRs before they reached reviewers~\cite{Vasilescu2014, Cassee2020}, allowing integrators to focus on higher-level concerns such as design and architecture~\cite{Pham2013}.
Similarly, Vasilescu et al.~\cite{Vasilescu2015} observed that CI adoption led to the discovery of more issues, suggesting improved internal QA.

Despite the growing role of automation, human judgment remained central to code review.
Teams continued to evaluate aspects that automation could not address—such as architectural decisions, naming conventions, clarity, requirement adherence, and downstream impact. 
Studies highlighted the value of code reviews not only for defect detection but also for knowledge sharing and team coordination~\cite{Bacchelli2013,Baum2016}. 
At the time, only humans could provide nuanced feedback, such as discussing design alternatives or explaining the rationale behind implementation choices. 
Automation could flag a null pointer or run a test, but it could not suggest a simpler design.

This era demonstrated that continuous peer review, supported by automation, was both viable and beneficial.
Ubiquitous version control, mandatory and tool-driven peer review, and the integration of automated testing and static analysis into the review pipeline were key innovations. 
By the early 2010s, these practices laid the foundation for the next evolutionary step, where automation would shift from a supporting role to a central pillar of the code review process, augmenting human reviewers and enforcing quality gates with increasing rigor.

\subsection{Automation-Assisted Era}
\label{subsec:automation-assisted-era}
The Automation-Assisted Era (2010s-2020s) is characterized by a deepened partnership between human reviewers and automated tools, while building upon the Integrated Code Review Era.
Beginning in the early 2010s, automation evolved from a helpful supplement into a central organizing principle of the code review process. \textit{Optional} static analysis tools of the former era became an ordinary and integral part of the code review process across the industry.
Large technology companies integrated custom static analysis platforms directly into their review interfaces, with Google's Tricorder~\cite{Sadowski2015} and Facebook's Infer~\cite{Distefano2019} being prominent examples.

Continuous Integration and Continuous Delivery (CI/CD) pipelines became mandatory quality gates for code changes, and advanced tools, such as Machine Learning (ML)-based systems, began to assist or augment human reviewers in decision-making.
Earlier practice treated automation as optional support. Teams in this era increasingly treat automation as a co-reviewer that must sign off on code health before or alongside human approval~\cite{Sadowski2018}.

This co-working philosophy had measurable effects on review dynamics.
Rahman and Roy~\cite{Rahman2017b} found in open source projects that PRs with successful CI builds received quicker code reviews, while failed builds often stalled the process entirely.
On the other hand, Cassee et al.~\cite{Cassee2020} showed that CI adoption in projects brought a decrease in PR discussion comments in the review process. 
In other words, automation was handling issues that previously would have required explicit human discussion, freeing reviewers to focus on other concerns.

Despite advances, automation has not displaced human reviewers but it has \textit{specialized} them.
The valuable outcomes of code review remained fundamentally human-centric: discussing alternative approaches, understanding change rationale, educating team members, and maintaining collective awareness of system evolution~\cite{MacLeod2018,Kononenko2016}. 
Bacchelli and Bird's~\cite{Bacchelli2013} finding that reviewers spend much of their effort on knowledge transfer and design improvement alongside defect detection continued to hold true.

Even at Google, with extensive automation infrastructure, human reviewers remain indispensable for evaluating non-functional requirements and subjective quality aspects that CI cannot rate~\cite{Sadowski2018}.
However, with the recent rise of large language models (LLMs) and agentic AI, the workflows and tools defining this era, once considered the cutting edge, are now being recontextualized as \textit{traditional} rather than contemporary code review systems. We explore the details of these traditional code review systems in Section~\ref{sec:traditional-code-review-systems}.

In summary, the Automation-Assisted Era transformed code review into a human-automation partnership.
The outcome was a more efficient process in various aspects.
This era is now setting the stage for the next phase of code review, currently unfolding in the 2020s, in which AI agents participate alongside human experts.
This shift represents a natural evolution of the ideas seeded in the Automation-Assisted Era, now maturing into a review process where AI agents act as collaborators rather than supporting tools.

\begin{table}[t]
\centering
\caption{High-level evolution of code review practices and their defining characteristics.}
\label{tab:code-review-evolution}
{\footnotesize
\renewcommand{\arraystretch}{1.15}
\begin{tabularx}{\columnwidth}{
>{\RaggedRight\arraybackslash}p{0.13\columnwidth}
>{\RaggedRight\arraybackslash}p{0.18\columnwidth}
>{\RaggedRight\arraybackslash}p{0.18\columnwidth}
>{\RaggedRight\arraybackslash}p{0.20\columnwidth}
>{\RaggedRight\arraybackslash}X
}
\toprule
\textbf{Era} &
\textbf{Review Practice} &
\textbf{Coordination Mode} &
\textbf{Quality Control \& Integration Policy} &
\textbf{Defining Characteristic \& Practice Shift} \\
\midrule
\textbf{Ad Hoc Code Review Era}\newline (1940s--1960s) &
Informal, reactive consultation &
Verbal feedback; handwritten margin notes &
No formalized roles; no structured processes; no approval gates &
Code review was not yet a systematic QA practice \\
\midrule
\textbf{Formal Inspection Era}\newline (1970s--1990s) &
Synchronous, structured, role-based inspection &
Moderator-led face-to-face meetings; checklists; defect logs &
Rework verified before artifacts could pass &
Review became a repeatable, measurable QA methodology \\
\midrule
\textbf{Lightweight Peer Review Era}\newline (1990s--2000s) &
Asynchronous, patch-based peer review &
Mailing-list discussion; quoted diffs; iterative replies &
Maintainer judgment and community consensus &
Review traded heavyweight inspection for velocity and accessibility \\
\midrule
\textbf{Integrated Code Review Era}\newline (2000s--2010s) &
Asynchronous, tool-supported peer review &
PRs; web/IDE tools; inline comments; reviewer assignment &
Explicit approval before merge; early CI checks &
Review became a scalable, platform-mediated collaboration mechanism \\
\midrule
\textbf{Automation-Assisted Era}\newline (2010s--2020s) &
Asynchronous, automation-assisted peer review &
PR platforms; CI/CD; bots; static analysis; automated signals &
Layered human approval and automated quality gates &
Automation became a routine CI/CD companion to code review, as human judgment remained the core of asynchronous review \\
\bottomrule
\end{tabularx}
}
\end{table}

\section{Traditional Code Review Systems}
\label{sec:traditional-code-review-systems}

This section presents a generalized workflow model for traditional code review systems (in the Automation-Assisted Era), illustrating the lifecycle from issue creation through PR resolution. Subsequently, the inherent limitations and bottlenecks associated with these manual workflows are discussed in Section~\ref{subsec:challenges-in-current-code-review-systems}.

\begin{figure}[h]
    \centering
    \includegraphics[width=\linewidth]{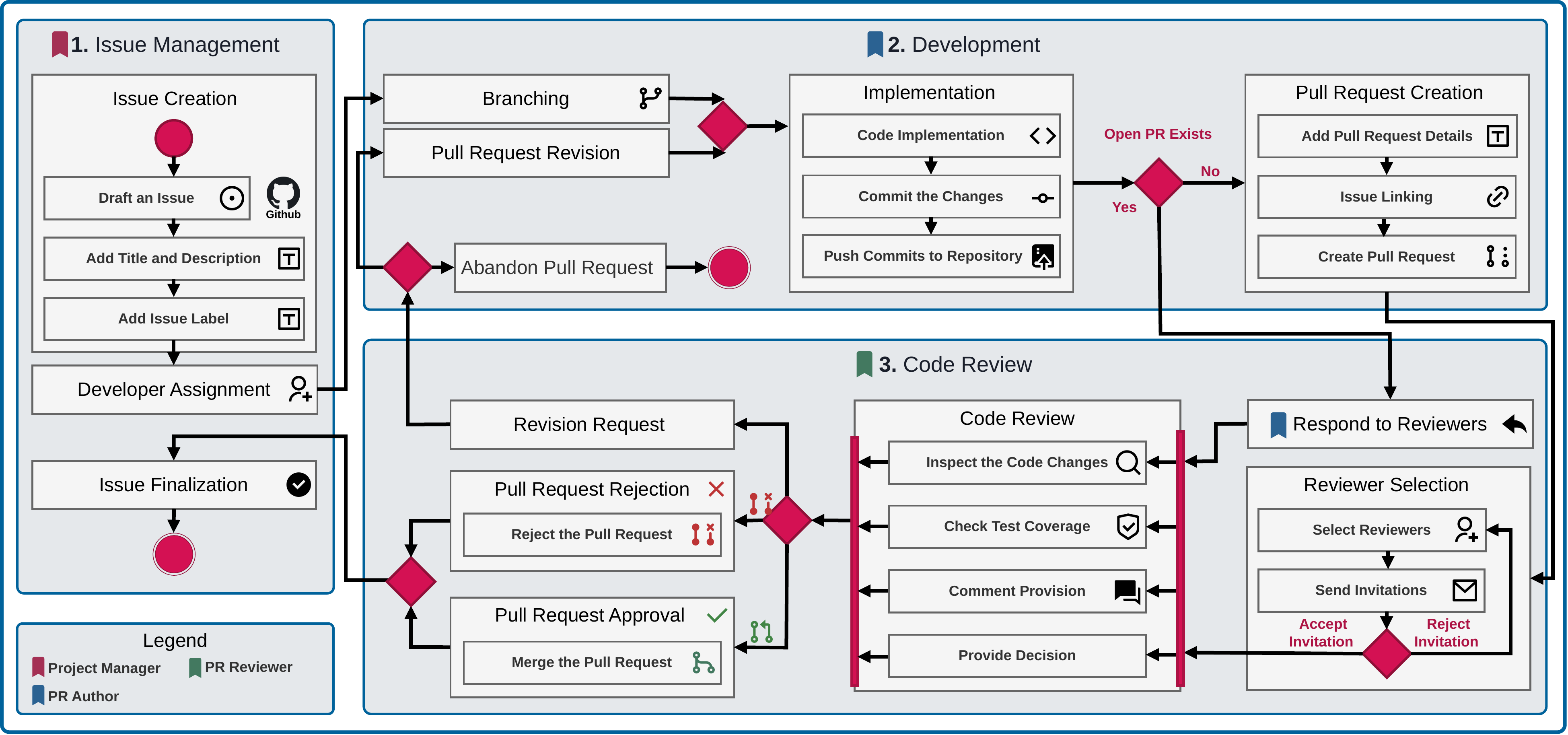}
    \caption{Overview of the traditional code review systems workflow.}
    \label{fig:current_code_review_systems}
\end{figure}

Traditional code review platforms function as unified ecosystems that consolidate VCSs, code review platforms, and issue management utilities into a single platform. Figure~\ref{fig:current_code_review_systems} illustrates this PR-based software development lifecycle, detailing the workflow from initial issue creation to the final decision regarding the acceptance or rejection of a PR involving multiple stakeholders. 

The workflow typically begins with a Project Manager, Product Manager, Team Lead, Developer, or Maintainer establishing a unit of work within issue management applications such as Jira or GitHub Projects, as shown in the \textit{Issue Management} stage in Figure~\ref{fig:current_code_review_systems}. Furthermore, unlike an industrial development environment, open source settings allow any community member, active user, or technical contributor to open an issue to report a bug or propose a new feature. This process involves defining the scope of the problem or requirement, often summarized in a concise sentence, followed by the addition of a specific title and detailed descriptive fields. These fields may vary based on the type of issue being created. While some issues are intended to fix bugs, others are intended to implement features, and these types are often distinguished through specific labels. Subsequently, the Project Manager or Team Lead determines the trajectory of the issue based on the project management methodology employed. This determination may lead to the direct assignment of the task to a developer or its placement into a project backlog. Within the backlog, issues are often organized using prioritization mechanisms such as severity labels, milestone targets, or urgency rankings which guide the selection process before a contributor or maintainer claims the task for implementation.

Following the assignment of an issue, the PR author starts development by creating a dedicated branch appropriate for the task, such as a feature branch for new functionality or a bug fix branch for defect resolution. The PR author then executes the necessary code modifications, ensuring that each logical unit of change is encapsulated within an atomic commit to maintain a clean history. Following the local completion of these changes, the developer pushes the code to the remote repository. To formally propose merging these modifications into the target branch, the PR author opens a PR utilizing Git hosting platforms such as GitHub or Bitbucket. During this submission process, the PR author establishes traceability by explicitly linking the PR to the corresponding issue and populating the PR metadata with a descriptive title and a summary of the implementation as depicted in the \textit{Development} stage in Figure~\ref{fig:current_code_review_systems}.
In subsequent iterations following reviewer feedback, the PR remains open; the PR author addresses the requested changes by committing additional modifications to the same branch, which are automatically reflected in the existing PR without requiring resubmission.

Once the PR is submitted, the lifecycle transitions to the \textit{Code Review} stage.
Here, responsibility for selecting reviewers and requesting their review depends on the software development environment, falling to Project Managers or PR authors. PR reviewers who accept the review request conduct a rigorous examination of the submitted PR, evaluating various aspects including code quality, functional correctness, and design adherence. This assessment generates feedback in the form of specific suggestions, clarification questions, or inline comments. This stage is illustrated in the \textit{Code Review} stage in Figure~\ref{fig:current_code_review_systems}. Based on this analysis, the workflow diverges into three potential resolutions.
 \begin{enumerate}
     \item If the contribution meets the project standards, PR reviewers approve the PR, authorizing the merge of the PR author's branch into the target codebase and typically concluding the workflow by closing the associated issue.
     \item PR reviewers may reject the PR if the changes are deemed unsuitable; in this scenario, the developer abandons the branch, while the underlying issue is either closed or kept open depending on the project's future needs.
     \item PR reviewers may request revisions, triggering an iterative cycle. In this case, the developer should either abandon the PR or address the reviews by implementing the requested changes, committing them, and pushing updates to the branch to solicit a subsequent round of review.     
 \end{enumerate}

The Code Review Workflow discussed in this section is intended to represent the code review process in its most generalized form, capturing the essential interactions common to today’s software engineering practices. However, it is important to acknowledge that specific practices, tooling configurations, and procedural mandates vary considerably across different organizations. For example, while some high-velocity teams may adopt lightweight, trunk-based development workflows with minimal blocking gates, organizations in regulated domains often enforce rigorous, multi-tiered approval hierarchies involving distinct roles such as security auditors or release managers. Zhang et al.~\cite{zhang2023} demonstrate that various influencing factors including author characteristics, PR characteristics, project characteristics, and tools may influence the lifecycle of the PR. Despite these operational divergences, the fundamental principles of PR-based software development, branch-based isolation, and asynchronous peer feedback remain broadly applicable across traditional code review systems.

\subsection{Challenges in Traditional Code Review Systems}
\label{subsec:challenges-in-current-code-review-systems}

This section discusses the challenges in traditional code review systems.
The challenges discussed in this section represent the primary challenges that our proposed AI-powered framework aims to address, each corresponding to one or more of the five stages described in Section~\ref{sec:ai-powered-framework}; however, given the breadth of the software engineering QA literature, the challenges of traditional code review systems extend beyond the challenges covered here.
These challenges are analyzed across four thematic domains.
Challenges in PR Creation are discussed in Section~\ref{subsubsec:challenges-in-pr-creation}.
Challenges in Reviewer Assignment are outlined in Section~\ref{subsubsec:challenges-in-reviewer-assignment}.
Challenges in Code Review are examined in Section~\ref{subsubsec:challenges-in-pr-review}.
Lastly, challenges in Comment Provision are addressed in Section~\ref{subsubsec:challenges-in-comment-submission}.

\subsubsection{Challenges in PR Creation}

\label{subsubsec:challenges-in-pr-creation}

Once PR authors reach a sufficient level of implementation, they initiate a PR to merge changes into the target branch. However, during PR creation, PR authors often deviate from established practices, frequently omitting necessary details, neglecting documentation, or failing to link relevant issues. These suboptimal practices exacerbate challenges within the code review process and degrade software quality.

\textbf{Missing PR Context:} This challenge refers to the failure to include descriptive PR titles and descriptions, which are fundamentally intended to explain the changes, context, and rationale behind the code modifications. An empirical study by Liu et al.~\cite{Liu2019} on a dataset of 333,001 GitHub PRs revealed that 34\% of descriptions were empty, showing how prevalent this issue is.
This persists despite multiple studies validating that understanding code differences and the underlying rationale is critical for reviewers~\cite{Tao2012,MacLeod2018}. PRs lacking this necessary information increase cognitive load and confuse PR reviewers~\cite{Ebert2019,Chouchen2021}. This confusion often necessitates clarification requests, producing ``ping-pong'' communication~\cite{Doan2022}.
It also compels reviewers to submit ``LGTM'' (Looks Good To Me) reviews, a code review smell formally referred to as the LGTM smell~\cite{Doan2022,Gon2024}, and ultimately causes delays~\cite{Ebert2019}.

\textbf{Lack of Documentation:} PR authors often neglect to write documentation for their implementation. This omission creates documentation debt, which can ultimately compromise the repository's maintainability~\cite{Alves2016,Mendes2016}. By failing to record the operational logic or architectural decisions accompanying code changes, PR authors obscure the underlying design rationale. This forces future maintainers to undertake significant reverse-engineering efforts, thereby degrading the long-term quality of the software. Furthermore, the lack of documentation deprives PR reviewers of essential context regarding the PR author's intent. Without this guidance, reviewers struggle to distinguish between intended behavior and potential defects, leading to increased confusion~\cite{Baum2019}. This cognitive strain often triggers negative coping mechanisms, such as performing LGTM smells~\cite{Doan2022,Gon2024}. Alternatively, confused reviewers may ask numerous clarifying questions to understand the PR, thereby extending the code review cycle and causing delays~\cite{Chouchen2021,Ebert2019}.

\textbf{Missing Issue Links:} The omission of artifact traceability links is a detrimental practice affecting both code quality and review. Artifact traceability links are used to track requirements and the corresponding code changes that implement them~\cite{istreview}. In current software development, Issue-based Requirement Tracking (I-RT) is the predominant method~\cite{Lyu2023}. I-RT focuses on the ability to trace relationships from issue reports to other software artifacts, recognizing that in current workflows, issue reports often serve as the entry points for PRs; therefore, each PR should be linked to an issue. Beyond enhancing long-term repository maintainability and traceability by allowing maintainers to quickly understand context via tracing, these links are also useful for PR reviewers. Reviewing code without prior knowledge of the specific requirements or bug fixes can lead to suboptimal reviews~\cite{Ram2018,Doan2022}. However, due to a lack of enforced rules and oversight, developers often fail to link issues to their PRs~\cite{Rath2018}. Empirical studies underscore this issue: Bachmann et al.~\cite{Bachmann2010} found 52.4\% of bug-fixing commits unlinked, and Dogan et al.~\cite{Doan2022} reported missing links in 34.3\% of PRs. Such unlinked PRs reduce traceability and maintainability in the long term. On the code review side, the lack of requirements context can confuse reviewers, prompting requests for issue details that trigger ``ping-pong'' communication, or leading to LGTM smells~\cite{Doan2022,Gon2024} and subsequent delays. Therefore, ensuring the consistent maintenance of PR-issue links remains a persistent challenge.

\subsubsection{Challenges in Reviewer Assignment}
\label{subsubsec:challenges-in-reviewer-assignment}

Upon submission of the PR, the code review process begins with the assignment of PR reviewers. Depending on the specific workflow of the code review system, this is achieved through the author inviting peers, a project manager or team lead assignment, or an automated bot selecting candidates. To ensure an effective code review process, the selected PR reviewers should possess familiarity with the code changes, have adequate experience in code review, and not be burdened with excessive workload~\cite{Aurum2002,Rigby2011Paper}. However, in large-scale software teams, identifying the optimal reviewer by balancing expertise, experience, and availability while maintaining healthy team dynamics is often time-consuming and challenging~\cite{etin2021,Mashayekhi1993}.

\textbf{Finding the Right Reviewer:} A notable difficulty lies in identifying a reviewer with specific domain expertise; finding a reviewer familiar with the modified code chunk is essential for quality, as reviewers with more experience in a specific module are significantly less likely to miss defects~\cite{Kononenko2015}. Empirical studies demonstrate that familiarity with the code is one of the most influential factors determining code review duration and quality~\cite{Kononenko2015}. Alongside module expertise, finding reviewers with general experience in the review process itself presents another layer of difficulty. Reviewer experience is a key determinant of review quality and timeliness, as experienced reviewers tend to provide more useful feedback and navigate the process more efficiently~\cite{Rahman2017}. Although reviewer expertise fit reduces code comprehension effort, consistently assigning the optimal reviewer may not always be achievable in operational settings. In these scenarios where ideal matching fails, the proposed AI-powered framework provides critical technical context to available reviewers, directly strengthening the overall review process.

\textbf{Workload Distribution:} The equitable allocation and distribution of reviewer workload is a persistent challenge in reviewer selection. Workload is one of the most influential factors impacting code review quality and merge times~\cite{Kononenko2016}. In large software teams, accurately monitoring and balancing workload may be challenging. Suboptimal reviewer selection often results in specific experts accumulating excessive workloads~\cite{Ruangwan2019}. This imbalance leads to adverse outcomes: overloaded reviewers may reject review invitations, or if they accept, they may resort to LGTM smells to clear their review queue quickly. Doğan and Tüzün~\cite{Doan2022} explicitly identify reviewer availability as a direct root cause of the LGTM smell, characterizing it as situations in which the reviewer is too busy with other tasks but cannot decline the review request. Empirically, Gon et al.~\cite{Gon2024} report that 64.7\% of PRs across five large-scale projects are reviewed without any comment, with such comment-free reviews exhibiting the LGTM smell 3.5 times more frequently than commented reviews. Furthermore, Gon et al.'s~\cite{Gon2024} findings indicate that high workload contributes to sleeping reviews, where PRs stagnate without feedback, effectively stalling the development pipeline and extending code review turnaround times. Another challenge arises from the need to ensure equal and effective knowledge distribution during the review process. One of the primary motivations for code review is sharing knowledge among peers and maintaining team awareness~\cite{Bacchelli2013}. To exercise this practice, developers or managers may intentionally assign less experienced reviewers or those unfamiliar with a specific module to the PR~\cite{Ebert2019,Kovalenko2020}. While this fosters long-term benefits, it creates a tension with the immediate need for expert defect detection. Therefore, balancing the selection of experts for QA against the selection of novices for knowledge distribution and team awareness remains a persistent challenge in reviewer selection.

\textbf{Suboptimal Selection Practices:} The integrity of the PR reviewer selection process is often compromised by suboptimal practices by PR authors. In the absence of strict controls, authors may superficially assign the same set of reviewers repeatedly or assign themselves to review their own code to bypass rigorous code review~\cite{Doan2022}. Such practices significantly harm knowledge sharing. By reducing team awareness and avoiding critical feedback, these behaviors hinder the overall effectiveness of the code review process. Consequently, preventing such manipulation and ensuring objective PR reviewer selection becomes another operational challenge.

Altogether, selecting an optimal PR reviewer remains one of the most consequential challenges in traditional code review systems, as reviewer familiarity with the modified code is among the strongest predictors of defect detection rates~\cite{Kononenko2015}. The inability to consistently identify the optimal reviewer hinders the fundamental benefits of the practice, including QA, knowledge transfer, and team awareness, while simultaneously increasing developer workload and consequently causing significant process delays.

\subsubsection{Challenges in Code Review}

\label{subsubsec:challenges-in-pr-review}

Once the PR reviewer selection is complete and reviewers have accepted their invitations, the code review process commences. PR reviewers begin by inspecting the code changes, verifying test coverage, and analyzing the associated issue details. This process is arguably the most important component of code review workflows~\cite{Bacchelli2013,Sadowski2018}, serving as the primary gatekeeper for establishing QA, identifying defects, mitigating technical debt, and detecting security vulnerabilities before deployment.

\textbf{Change Understanding and Defect Detection:} A central difficulty in code review is change understanding, which refers to the necessity of comprehending the code change in minute detail and from multiple perspectives to verify correctness, design alignment, dependencies, and potential impacts. Although defect identification is one of the primary objectives of code review~\cite{Bacchelli2013}, effectively achieving this requires a deep understanding of the codebase, as subtle logical errors and edge cases are often indistinguishable during code review. Factors such as suboptimal PR reviewer selection, time pressure, and low code ownership can significantly impede defect detection~\cite{Bird2011,Matsumoto2010}. Empirical evidence from large-scale software ecosystems corroborates the complexity of this endeavor: a study by Kononenko et al.~\cite{Kononenko2015} on the Mozilla project revealed a significant gap in defect detection efficacy, reporting that 54\% of code reviews failed to identify bugs present in approved commits. Similarly, Czerwonka et al.~\cite{Czerwonka2015} highlighted the scarcity of defect-oriented feedback at Microsoft, finding that a mere 15\% of reviewer comments specifically pointed to potential defects.
Furthermore, the comprehension demand of code review is not uniform across feedback types.
Beller et al.~\cite{Beller2014} manually classified over 1{,}400 review-induced changes in two open-source projects and reported that 75\% of these changes are maintainability-related while 25\% address functional concerns, mirroring earlier industrial findings.
Maintainability-oriented comments such as naming, local readability, or documentation can often be produced from the diff itself, whereas functional-defect detection continues to require change-wide comprehension and behavioral reasoning.
These findings collectively underscore the substantial cognitive challenge and frequent fallibility inherent in manual defect identification tasks.

\textbf{Understanding Change Impacts:} Evaluating the broader, often ripple-like impact of a PR presents significant complexity beyond the immediate scope of the modified lines. During the code review process, PR reviewers may easily spot obvious syntactic or logical errors within the localized code diff. However, accurately predicting the side effects of those changes on distant project modules and ensuring system-wide consistency upon merge is significantly more challenging~\cite{Li2013}. This difficulty stems from the complex inter-dependencies and tight coupling often present in large-scale software architectures, where a modification in one component can inadvertently break functionality in another. This is especially crucial in mission-critical systems, where a single unexpected change can trigger severe consequences or regression defects in dependent modules that are not immediately visible in the diff~\cite{Fowler2004}. Change Impact Analysis (CIA) offers a methodological approach to estimate the potential effects of proposed changes and to detect hidden errors in dependent modules~\cite{Sun2010,Bohner1996}. Applying it manually, however, is rarely feasible within the time constraints of an active PR review. Performing such analysis manually requires PR reviewers to construct a complex mental model of the entire software system's execution flow, which is often cognitively overwhelming or impossible given time constraints. Compounding this difficulty, traditional code review systems exacerbate this issue by failing to explicitly visualize the potential execution impact or call-graph dependencies of proposed changes, leaving PR reviewers to rely on intuition and incomplete knowledge~\cite{Blischak2016}.

\textbf{Understanding PR-Issue Alignment:} Determining whether a PR completely and accurately implements the requirements specified in the issue is a critical challenge for PR reviewers. A recent study by Isik et al.~\cite{Isik2025} formalizes this concept as PR-issue alignment, defining four alignment categories: Exact (PR fully addresses requirements without unrelated changes), Tangling (PR includes unrelated changes), Missing (PR fails to fully address the issue), and Missing and Tangling (combining both deviations). If not adequately addressed, Missing PRs serve as indicative signs of technical debt~\cite{Tom2013,Alves2016}. Conversely, Tangling PRs hinder code review effectiveness and significantly impede defect detection; the inclusion of irrelevant tasks creates noise, increasing the risk that reviewers will miss critical changes hidden within the unrelated code~\cite{Herzig2016,Tao2015}. Tangling PRs may also cause delayed reviews because, although the majority of a PR might be correct, a small, controversial, or unrelated part can block the approval of the entire PR~\cite{Pascarella2018}. Multiple studies highlight the prevalence of this issue; Herzig et al.~\cite{Herzig2016} show that 7-20\% of changesets contain tangling commits, supported by further prevalence studies~\cite{Herzig2013,Herbold2022}. Additionally, Isik et al.~\cite{Isik2025} found that 16.5\% of PRs were labeled as Missing, demonstrating the severity and importance of this challenge in maintaining software quality.

\textbf{Large Code Diffs:} PRs with large code diffs impose a significant review challenge. While best practices dictate that PR authors should create concise, atomic PRs that address a single concern to facilitate easier code review~\cite{Rigby2013}, practical constraints arising from task complexity or poor development habits often lead to PRs with extensive code diffs~\cite{Doan2022}. Reviewing these large PRs presents a substantial challenge~\cite{MacLeod2018,Sadowski2018}, as they impose a heavy cognitive load that can easily overwhelm and confuse reviewers~\cite{Ebert2019,Baum2019}. This increased cognitive load negatively impacts the effectiveness of the review process, the quality of code review comments, and turnaround time~\cite{Ebert2019,Baum2019}. Moreover, PRs with large code diffs reduce PR reviewers' ability to detect defects~\cite{Czerwonka2015} and result in less useful comments~\cite{Bosu2015}. Furthermore, they typically require more revisions than smaller changes and face a higher risk of abandonment~\cite{Wang2019}. Ultimately, these large changesets delay code review~\cite{Baysal2016}, demonstrating the difficulty of reviewing such contributions manually and necessitating tools to aid PR reviewers' understanding.

\textbf{Time Pressure:} Time constraints and pressure serve as a critical barrier to effective code review~\cite{Kononenko2016,MacLeod2018}. Although timely feedback is essential for developers, best practices advise against rushing; studies suggest that code should not be reviewed faster than 200 lines per hour to maintain inspection quality~\cite{MacLeod2018,Kemerer2009}. However, organizational realities often conflict with this standard. Strict shipping deadlines may force developers to prioritize coding over reviewing, generating significant time pressure~\cite{Ruangwan2019}. Furthermore, suboptimal workload distribution often leads to excessive review queues for certain team members~\cite{Baysal2013}. This workload pressure can degrade QA and induce negative process behaviors. Under increased time pressure, PR reviewers may inspect changes hastily, increasing the likelihood of missed defects~\cite{Kemerer2009,McIntosh2016}, performing LGTM smells~\cite{Doan2022,Gon2024}, or leaving PRs unreviewed~\cite{Wang2019}. Additionally, long review queues directly cause delays in code review~\cite{Baysal2013}. These findings show the challenge of balancing rigorous review standards with the need for efficiency in fast-paced environments.

\subsubsection{Challenges in Comment Provision}

\label{subsubsec:challenges-in-comment-submission}

Following the completion of the code review, PR reviewers provide feedback, suggestions, and critique through review comments. These comments can be either anchored to specific lines within the code diff or submitted as general remarks regarding the overall PR. The quality and nature of these comments are pivotal, directly influencing code review process effectiveness, knowledge sharing, and delays.

\textbf{Comment Usefulness:} To ensure an effective code review process, PR reviewers are expected to provide feedback that is useful, clear, informative, relevant, and polite. Useful comments are defined as those that constructively help the developer improve the PR~\cite{Bosu2015}. Specifically, a useful comment should identify defects, enhance code quality, utilize appropriate language, improve maintainability, or facilitate better design decisions~\cite{Turzo2024}. However, challenges such as large changesets or lack of experience often cause PR reviewers to submit non-useful comments that focus on trivial issues, such as coding style, while leaving deeper and more critical quality issues undiscussed~\cite{Bosu2015,Pangsakulyanont2014}. Bosu et al.~\cite{Bosu2015} demonstrated that 34.5\% of comments in Microsoft repositories were not useful, while Rahman et al.~\cite{Rahman2017} reported that 44.47\% of comments were classified as non-useful. The accumulation of such non-useful comments hinders the effectiveness of the process and extends merge times, making the provision of useful review comments a significant challenge.

\textbf{Sentiment and Toxicity:} The sentiment and tone of review comments are critical factors that influence the code review process beyond its technical aspects. The emotional content of feedback significantly impacts collaboration. Studies indicate that positive code review comments are associated with faster resolution times and strengthen team relationships~\cite{Asri2019}. Conversely, toxic or overly negative comments can have severe detrimental effects, causing mental health issues, increased stress, and burnout among developers~\cite{Sarker2023}, while simultaneously harming knowledge sharing and interpersonal relationships~\cite{Ahmed2017}. The prevalence of such toxic comments is non-negligible; an empirical study by Sarker et al.~\cite{Sarker2023} found that 19.1\% of comments in their dataset were labeled as toxic. Consequently, consistently maintaining a positive, encouraging and polite tone in code review comments constitutes a persistent challenge in traditional code review.

These challenges arise primarily from the intrinsic nature of the process~\cite{Bacchelli2013,MacLeod2018}, suboptimal practices employed by practitioners~\cite{Doan2022}, technical factors~\cite{Baysal2016}, and various human factors~\cite{Baysal2013,Ebert2019}.
If not adequately addressed, these issues may hinder the overall effectiveness of the code review process, cause process debt~\cite{Martini2019,Doan2022}, and prevent software development teams from realizing the full benefits of code review.
Therefore, it is crucial for the proposed AI-powered code review vision framework to thoroughly understand these challenges and develop mitigation strategies that target these root causes.

\section{AI-Powered Code Review Framework}
\label{sec:ai-powered-framework}

\begin{figure}[!htbp]
  \centering
    \includegraphics[width=\linewidth]{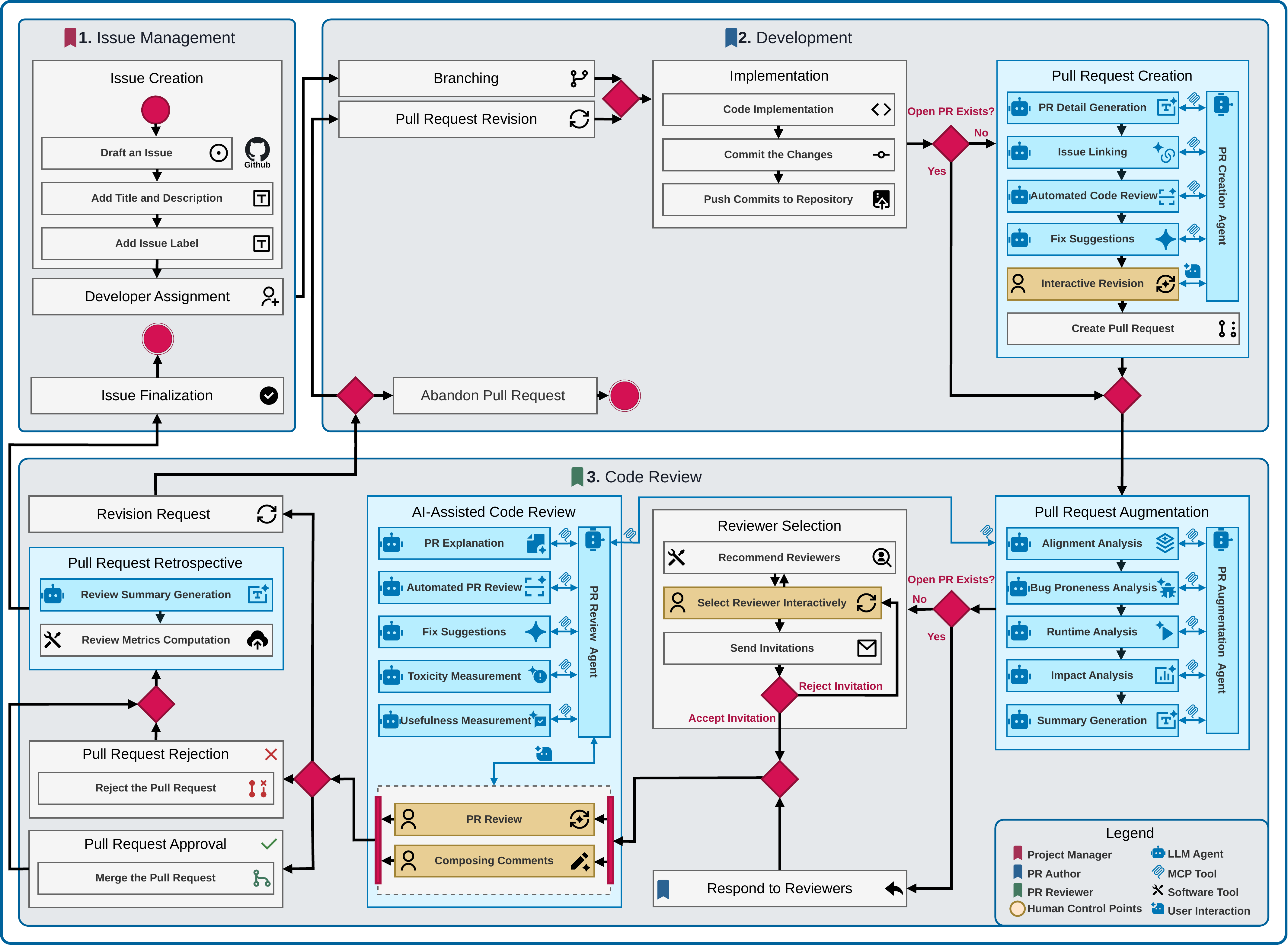}
  \caption{Overview of AI-Powered Code Review Framework Workflow}
\label{fig:ai_code_review_framework_workflow}
\end{figure}

To mitigate the challenges and suboptimal practices prevalent in traditional code review systems detailed in Section~\ref{subsec:challenges-in-current-code-review-systems}, we propose an AI-powered code review framework. By integrating LLM agents and traditional tools into the code review workflow, we aim to address existing limitations and enhance PR creation, reviewer selection, code review, and comment provision, as shown in Figure~\ref{fig:ai_code_review_framework_workflow}. This approach is designed to improve the effectiveness of the code review process by addressing existing procedural limitations. Alongside the integration of LLMs, we established explicit human-in-the-loop (HITL) quality gates to minimize automation bias, accumulated errors, and model hallucinations. By enforcing these manual verification points, the framework aims to achieve higher code quality, accelerated review cycles, and reduced cognitive fatigue for developers. Within our framework, the primary focus spans from the initial PR creation through the final approval or rejection of the PR, concluding with the PR Retrospective stage. The following subsections delineate the operational stages of the proposed framework. Section~\ref{subsec:overview} provides a conceptual overview at a high level while Section~\ref{subsec:pr-creation} describes the \textit{PR Creation} stage. Section~\ref{subsec:pr-augmentation} details the \textit{PR Augmentation} stage and Section~\ref{subsec:reviewer-selection} presents the \textit{Reviewer Selection} stage. Finally, Section~\ref{subsec:ai-assisted-code-review} explains the \textit{AI-Assisted Code Review} stage and Section~\ref{subsec:pr-retrospective} examines the \textit{PR Retrospective} stage.

\subsection{Overview}
\label{subsec:overview}

Once a PR author is assigned to an issue, they implement the necessary code changes locally. After the code modifications reach a sufficient level of maturity, PR authors initiate the process by creating a draft, as illustrated in the \textit{PR Creation} stage in Figure~\ref{fig:ai_code_review_framework_workflow}. At this stage, first, the \textit{PR Detail Generation Agent} automatically generates the PR title and description based on the provided code diff. Following this generation, the \textit{Issue Linking Agent} identifies relevant issues and establishes explicit traceability links, or it generates a new issue if no appropriate issue exists to link with the PR. Subsequently, an \textit{Automated Code Review Agent} inspects the proposed code diff to identify syntax errors or policy violations.
It then provides actionable review comments within the PR draft before the submission undergoes a more rigorous review process. Instead of merely detecting problems, the \textit{Fix Suggestion Agent} suggests concrete patches or minimal change snippets that resolve the identified issues, providing explicit statements regarding what the patch modifies and what it leaves intact. Following these automated steps, the PR author uses natural language to interactively revise the draft in collaboration with the \textit{PR Creation Agent} before requesting a formal human review. During this interactive phase, PR authors communicate with the agent to refine implementation details, adjust issue links, request additional automated checks, and apply suggested code repairs. This mandatory human verification step establishes a quality gate where the author confirms the validity of generated content to prevent silent drift in the linked issue context and PR details. The PR author finalizes and creates the PR only when all generated artifacts accurately reflect the intended implementation logic.

Once the PR is opened or updated, the framework proceeds with the \textit{PR Augmentation} stage illustrated in Figure~\ref{fig:ai_code_review_framework_workflow}. In this stage, specialized agents analyze four different dimensions of the PR to establish the analytical evidence required for the \textit{Reviewer Selection} and \textit{AI-Assisted Code Review} stages. The \textit{Alignment Analysis Agent} classifies requirement fulfillment into the exact, tangling, missing, or missing and tangling categories introduced by Isik et al.~\cite{Isik2025}. The \textit{Bug Proneness Analysis Agent} calculates risk scores utilizing historical defect density, code churn, and hotspot files to predict potential failures~\cite{hoang2019deepjit}. The \textit{Runtime Analysis Agent} executes the code in a sandbox environment to collect reproducible evidence including logs and execution traces. The \textit{Impact Analysis Agent} evaluates cross module dependencies to determine downstream effects on application interfaces by performing CIA~\cite{Jayasuriya2024}.
These four analysis agents operate concurrently, as each requires the code diff, the linked issue context, and agent-specific evidence as input.
After these four agents complete their tasks, the \textit{Summary Generation Agent} synthesizes their outputs into a structured summary containing verifiable claims. The \textit{PR Augmentation Agent} governs these subordinate agents and enables PR reviewers to ask targeted questions regarding any specific analysis dimension during the \textit{AI-Assisted Code Review} stage.

Following the \textit{PR Augmentation} stage, the workflow proceeds to the \textit{Reviewer Selection} stage illustrated in Figure~\ref{fig:ai_code_review_framework_workflow} when no reviewer has been assigned to the PR. During this stage, the framework utilizes traditional reviewer recommendation tools to identify optimal candidates by evaluating multiple quantitative factors including reviewer expertise, historical review experience, current workload queue, and familiarity with the code changes. Based on these metrics, the PR author selects the appropriate PR reviewers from the recommendation list to send invitations. After the selected reviewers accept their invitations, the workflow continues with the \textit{AI-Assisted Code Review} stage.

Once PR reviewers begin the PR review, the framework assists them through a \textit{PR Review Agent} that orchestrates multiple specialized agents or tools within the \textit{AI-Assisted Code Review} stage as illustrated in Figure~\ref{fig:ai_code_review_framework_workflow}. PR reviewers communicate with this agent utilizing natural language. To establish a shared base for human PR reviewers and agents, the proposed system generates a semantic \textit{diff-map}, a representation of the code diff organized around logical units such as functions and classes rather than sequential file differences, with anchors and explicit traceability links to prior analytical reports attached to each unit; Section~\ref{subsec:ai-assisted-code-review} elaborates on this representation. This representation transforms the review process from a memory exercise into a verifiable retrieval task, verifiable in the provenance sense: every reviewer-visible claim is anchored to its source-code location and to the report that produced it. Through the \textit{PR Review Agent}, PR reviewers access the \textit{Explanation Agent} to help them understand the code by answering questions about the PR. Concurrently, the \textit{Automated PR Review Agent} reviews the modifications, and the \textit{Fix Suggestion Agent} proposes actionable repairs that are forwarded to the PR author as suggestions upon PR reviewer approval. Furthermore, the system integrates enhanced comment submission features accessible via the review menu. A \textit{Toxicity Measurement Agent} evaluates the toxicity of review comments while reviewers compose their feedback, since toxic comments deter future contributions and erode team collaboration~\cite{Sarker2023}, while a \textit{Usefulness Measurement Agent} identifies superficial remarks to mitigate potential bikeshedding~\cite{Turzo2024}. PR reviewers can also ask the \textit{PR Review Agent} to retrieve existing insights or request additional details and executions from the \textit{PR Augmentation Agent}, such as fresh runtime traces or PR-issue alignment reevaluations, directly overlaid on the \textit{diff-map}. After finalizing their review, the PR reviewers decide to approve or request specific revisions before the framework moves to the \textit{PR Retrospective} stage.

Once the PR is approved or rejected after the \textit{AI-Assisted Code Review} stage, the framework continues with the \textit{PR Retrospective} stage, illustrated in Figure~\ref{fig:ai_code_review_framework_workflow}, to capture PR review context knowledge. During this stage, the framework generates PR review summaries that capture review details and implementation summaries to preserve repository memory. Furthermore, the framework computes metrics and collects data regarding the PR review process to facilitate continuous process improvement. By synthesizing all socio-technical data points captured throughout the entire workflow, the framework ensures thorough documentation of the PR lifecycle before the final closure of the issue and the conclusion of the PR.

\subsection{PR Creation}
\label{subsec:pr-creation}

In modern software development, developers use PRs to propose code changes. When submitting PRs on VCS platforms like GitHub or GitLab, authors must manually specify branches, write a title and description, and assign labels. However, this manual process frequently leads to incomplete details. For instance, Liu et al.~\cite{Liu2019} found that 34\% of PR descriptions are entirely empty, which increases PR reviewer cognitive load and delays integration. Similarly, Dogan et al.~\cite{Doan2022} report that 34.3\% of PRs lack traceability links to issues, creating missing context that consumes excessive review time and harms long-term maintainability. To address these inefficiencies, our AI-powered code review framework autonomously generates PR details, establishes traceability links, provides initial review comments, and suggests code repairs. Consequently, PR authors verify and interactively refine these outputs with the \textit{PR Creation Agent} before formally submitting the PR as illustrated in the \textit{PR Creation} stage of Figure~\ref{fig:ai_code_review_framework_workflow}. If a PR is already open and the PR author is responding to review comments, our framework bypasses the \textit{PR Creation} stage and proceeds directly to the \textit{PR Augmentation} stage.

\subsubsection{PR Detail Generation} 
\label{subsubsec:pr-detail-generation}
PR details, comprising the title and the description, constitute an essential part of the code review process because PR reviewers depend on these artifacts to understand the rationale and scope of the proposed code modifications. Traditionally, PR authors write the PR title and description manually before requesting review. However, recent academic research demonstrates a distinct shift toward automating this process. For instance, Liu et al.~\cite{Liu2019} propose an automated approach that treats description generation as a text summarization problem by leveraging commit messages and source code comments. Then, Zhang et al.~\cite{Zhang2022} introduce the specific task of automatic PR title generation by formulating it as a one sentence summarization challenge. Hu et al.~\cite{Hu2022} investigate the evaluation metrics of such generated text, demonstrating the necessity of correlating automated scoring mechanisms with human assessments to ensure documentation quality. Irsan et al.~\cite{irsan2022autoprtitle} further operationalize this concept by introducing \textit{AutoPRTitle}, a tool that utilizes a fine-tuned bidirectional and auto regressive transformer model to generate precise PR titles. Similarly, Sakib et al.~\cite{Sakib2024} demonstrate that fine-tuning a text to text transfer transformer model on large PR datasets significantly outperforms baseline summarization algorithms. Reflecting this academic trajectory, commercial VCS platforms have begun integrating similar capabilities, as evidenced by \textit{GitHub Copilot}\footnote{\url{https://github.com/features/copilot}}, \textit{GitLab Duo}\footnote{\url{https://about.gitlab.com/gitlab-duo/}}, and \textit{Bitbucket AI}\footnote{\url{https://www.atlassian.com/software/bitbucket/features/ai}}, which directly represent the practical application of our proposed vision.

Building upon these foundational studies and commercial advancements, our proposed framework utilizes a \textit{PR Detail Generation Agent} that employs LLMs specifically fine-tuned on VCS histories to accurately summarize code differences. The PR Detail Generation Agent automatically constructs the title and the description the moment a developer begins drafting the PR within the platform. The PR authors review the generated PR details and execute iterative revisions using natural language commands to revise PR details. For example, if a PR author needs to modify a specific parameter name within the generated description, they select the relevant text section, which triggers a localized chat interface. The PR author then types the specific modification request into the chatbox, and the proposed system fulfills the request by delegating the contextual refinement task back to the PR Detail Generation Agent.

\subsubsection{Issue Linking} 
\label{subsubsec:issue-linking}
Once PR details are generated, our AI-powered code review framework proceeds to establish relevant issue links. Traditionally, developers manually link issues using keywords such as ``fixes'' or ``resolves''. While these connections are critical for repository traceability and providing reviewers with functional context, developers frequently omit them. Bachmann et al.~\cite{Bachmann2010} show that up to 52.4\% of bug fixing commits lack these necessary links. To address this gap, researchers have explored various automated linking mechanisms. Li et al.~\cite{li2018linking} define these connections as an Issue Unit Network, demonstrating that while critical for identifying complex dependencies, they are frequently omitted. To automate recovery, Partachi et al.~\cite{partachi2022aide} introduced \textit{Aide memoire}, an ML tool that recovers missing links with high precision to preserve project memory. The traceability benefits of such issue linking tools are validated by the industrial evaluation of the \textit{ReLink} tool by Yasa et al.~\cite{yasa2024relink}, which shows that practitioners prioritize explainable systems that provide explicit confidence scores. Furthermore, Pilone et al.~\cite{pilone2025augmenting} demonstrated that LLMs can accurately map internal GitHub issues to external user reviews, providing a richer context beyond technical specifications.

Building upon these studies, our framework utilizes an \textit{Issue Linking Agent} employing LLMs similar to~\cite{pilone2025augmenting} to autonomously link the optimal matching issue. The agent extracts semantic keywords and embeddings from the code diff, commit messages, branch name, PR title, and PR description to generate LLM-optimized search queries. Using a Retrieval-Augmented Generation (RAG) module, it queries a Vector Database via a hybrid approach combining keyword matching with semantic search. The agent then performs cross-encoder re-ranking on retrieved candidates to prioritize contextually relevant results based on the technical intent of the PR. If the top-ranked result satisfies the configured relevance threshold, the agent links the most relevant issue to the PR; when the PR addresses multiple requirements, all issues whose relevance scores exceed the threshold are linked. If no result satisfies the configured threshold, the agent uses text generation utilities of LLMs to create a new issue with a relevant title and description. Finally, authors verify and interactively modify these traceability links before submission.

\subsubsection{Automated Code Review} 
\label{subsubsec:automated-code-review}
Once the relevant issue is linked to the PR draft and the necessary connection is established, the framework continues to the review of the PR. Manual code review requires human reviewers to carefully read proposed modifications line by line to identify common problems such as inconsistent error handling, missing edge cases, unsafe Application Programming Interface (API) usage, and local convention violations. While this rigorous human evaluation is necessary, PR authors must proactively ensure that their submissions do not contain trivial defects or stylistic inconsistencies often referred to as bikeshedding before requesting a review. Additionally, receiving timely feedback presents a significant challenge in the development lifecycle, demonstrating the value of automated code review in providing instant feedback. Because this manual inspection is highly costly, recent academic research has focused on LLMs to automate code review. For example, Li et al.~\cite{li2022codereviewer} introduced \textit{CodeReviewer}, which utilizes large scale pre-training specifically targeting code review activities. However, because generative models occasionally produce vague suggestions, subsequent research by Li et al.~\cite{li2025noisy} demonstrated that filtering dataset noise is essential for producing high signal feedback. Furthermore, Zhou et al.~\cite{hong2024commentfinder} proposed \textit{CommentFinder}, a retrieval based approach that recommends comments resembling prior human feedback, while Jaoua et al.~\cite{jaoua2025combining} combined LLMs with static analyzers to ensure generated proposals are grounded in concrete code issues. This agentic approach extends to specialized domains, evidenced by Chen et al.~\cite{chen2024security} developing a multi-agent system for security vulnerability identification. As these systems increase in complexity, researchers have established dedicated evaluation frameworks such as \textit{EvaCRC}~\cite{Yang2023EvaCRC}, \textit{DeepCRCEval}~\cite{springer2025deepcrc}, and \textit{CRScore}~\cite{naacl2025crscore}, alongside comprehensive benchmarks~\cite{zhang2024benchmarking}.
Beyond academic prototypes, tools like Qodo PR Agent~\cite{qodo2025}, CodeRabbit~\cite{coderabbit2025}, and GitHub Copilot code review~\cite{githubCopilotCodeReview} are now deployed in industrial environments.
Despite empirical evaluations by Cihan et al.~\cite{Cihan2025} and Sun et al.~\cite{sun2025bitsai} demonstrating the practical utility of these tools, significant challenges remain. Watanabe et al.~\cite{watanabe2024chatgpt} highlight mixed developer trust, while other studies emphasize the risks of hallucinations~\cite{Tufano2025}, context switching friction~\cite{esem2025rethinking}, and systemic failures in agentic coding ecosystems~\cite{arxiv2026fail, arxiv2025agentic}.

Building upon this research and addressing the highlighted limitations, our framework introduces a robust automated review pipeline driven by interactive collaboration. Following the automatic generation of PR details and issue links, an \textit{Automated Code Review Agent} retrieves issue and PR details to identify syntax errors and policy violations before the PR undergoes a rigorous human review process. To maximize developer trust and mitigate the risk of hallucinations, this agent formulates findings as actionable proposals rather than absolute verdicts, tying each review item to a concrete location within the diff alongside a brief rationale. To facilitate quick resolution, the framework subsequently utilizes a \textit{Fix Suggestion Agent} to generate patches for the identified issues. Finally, the PR author interactively revises the PR draft using natural language when they inspect the automated findings. During this iterative phase, PR authors interact with the system to inquire about identified rationales and request additional review cycles for modified logic.

\subsubsection{Fix Suggestion} 
\label{subsubsec:pr-fix-suggestion}
Once the \textit{Automated Code Review Agent} comments on the PR, our AI-powered code review framework advances to the Fix Suggestion phase. Transitioning from detecting defects to automatically resolving them addresses the implementation friction caused by traditional manual repair. To provide actionable resolutions, early industrial systems like \textit{SapFix}~\cite{marginean2019sapfix} automatically generated production patches at Meta, while \textit{Getafix}~\cite{bader2019getafix} learned from past human edits to suggest natural fix patterns. More recently, research targeting the code review interface introduced \textit{AutoTransform}~\cite{thongtanunam2022autotransform} to automate tedious review modifications, and Frommgen et al.~\cite{frommgen2024resolving} demonstrated using ML to resolve natural language comments with concrete edits. However, effective automated repair requires developer trust and auditability, necessitating clear signals~\cite{petrovic2023mutant} and standard review transparency~\cite{endres2022repair}. In practice, this resolution via patch model is rapidly being adopted. Utilities like \textit{reviewdog}~\cite{reviewdog} convert static analysis into actionable PR comments, and this paradigm is evolving into fully autonomous agents. 
Tools like \textit{GitHub Copilot CLI}~\cite{githubCopilotCLI}, \textit{Cursor}~\cite{cursor}, \textit{Claude Code}~\cite{claudecode}, and \textit{Windsurf}~\cite{windsurf} represent this frontier by generating real-time patches that transition the author from manual implementation to supervisory verification.

Building upon these studies and tools that generate patches to improve overall quality, our AI-powered code review framework operationalizes these capabilities directly within the \textit{PR Creation} stage to enhance the submission before formal review. The \textit{Fix Suggestion Agent} produces patch suggestions for comments left by the \textit{Automated Code Review Agent}. This agent provides explicit statements regarding exactly what the patch modifies and what structural logic it leaves intact, making the review items highly comparable and auditable. These suggestions appear directly on the code diff as actionable patches that the PR author can approve or reject. Following these, the PR author utilizes natural language to interactively refine the PR draft and request supplementary code repairs.

\subsubsection{Interactive Revision} 
\label{subsubsec:interactive-revision}
Once the \textit{PR Detail Generation Agent}, the \textit{Issue Linking Agent}, the \textit{Automated Code Review Agent}, and the \textit{Fix Suggestion Agent} complete their tasks, the PR author interactively verifies and revises the PR draft. The \textit{PR Creation Agent} uses these specialized agents as accessible tools to facilitate this collaborative step. Through natural language communication, the PR author directs the \textit{PR Creation Agent} to execute a wide variety of refinement commands. Developers can ask questions about, edit, or modify the generated PR title and description. Furthermore, the PR author can request the system to search for alternative candidate issues or explicitly link the draft to a different issue within the repository. As the implementation evolves, developers can ask for another round of automated code review to evaluate newly added logic. Crucially, the developer can select and directly apply the concrete patches provided by the \textit{Fix Suggestion Agent}, or they can ask the system to generate alternative fix suggestions to explore different remediation strategies. By enforcing this manual human verification process, our AI-powered code review framework ensures verifiability, higher accuracy, alignment with the original intent, and traceability while preserving ultimate PR author authority over the software repository. The PR author finalizes and creates the PR only when all generated artifacts accurately reflect the intended implementation logic. Once the PR author formally submits this verified PR draft, our AI-powered framework continues to the \textit{PR Augmentation} stage to establish the evidence and rigorous analytical foundation for both the Reviewer Selection stage and the AI-Assisted Code Review stage.

\subsection{PR Augmentation}
\label{subsec:pr-augmentation}

Following PR creation or update, our visionary code review system continues with the \textit{PR Augmentation} stage. In this stage, specialized agents evaluate PR-issue alignment, risk profile, change impacts, and runtime behavior. These insights provide technical factors for the \textit{Reviewer Selection} stage and assist PR reviewers during the \textit{AI-Assisted Code Review} stage. In traditional code review systems, understanding the PR remains a major challenge~\cite{Bacchelli2013, MacLeod2018, Davila2021}. Inadequate understanding prevents PR reviewers from verifying correctness, dependencies, and potential impacts, and it causes them to miss defects. Understanding cross module side effects to ensure system wide consistency is particularly difficult~\cite{Li2013}. Furthermore, verifying PR-issue alignment may be challenging.
Empirical evidence shows that 7 to 20\% of changesets contain unrelated tangling commits~\cite{Herzig2016} and that 16.5\% of changesets fail to fully address the intended issue~\cite{Isik2025}. Additionally, large code diffs impose a heavy cognitive load that confuses PR reviewers and degrades review effectiveness~\cite{Czerwonka2015}. To mitigate integration delays and missed defects while assisting PR reviewers in achieving a rigorous understanding of complex modifications, our framework implements the \textit{PR Augmentation} stage to generate technical analytical foundations and critical evidentiary artifacts.

To resolve these challenges, the \textit{Alignment Analysis Agent} analyzes and categorizes PR-issue alignment into four distinct categories~\cite{Isik2025}. Simultaneously, the \textit{Bug Proneness Analysis Agent} calculates risk scores using historical defect density, code churn, and hotspot files to predict potential failures. The \textit{Impact Analysis Agent} performs CIA by evaluating cross-module dependencies to determine downstream effects on application interfaces and operational performance. The \textit{Runtime Analysis Agent} executes the code in a designated environment to collect reproducible evidence like logs, execution traces, and UI renderings. Finally, the \textit{Summary Generation Agent} summarizes these outputs into a structured summary containing verifiable claims. These analysis reports are subsequently added to the PR review panel. Each specialized agent is connected to the \textit{PR Augmentation Agent} as a tool, integrating distinct panels for PR-issue alignment, bug proneness, CIA, runtime analysis, and summary directly into the PR review panel. Each agent emits structured reports conforming to a common schema, enabling the \textit{Summary Generation Agent} to trace every synthesized statement to its originating analysis. Supporting our vision, major VCS providers began offering extensions for custom UI panels within PR review tabs. GitHub allows bots to submit analysis reports via PR comments. Bitbucket offers \textit{Forge}\footnote{\url{https://developer.atlassian.com/platform/forge/}}, a serverless application platform, to integrate custom UI panels directly into the PR review menu. Microsoft Azure Repos provides comparable UI extension capabilities for developers. These advancements demonstrate that current VCS platforms are actively preparing the infrastructure necessary to host the advanced plugins and integrated analysis capabilities of our visionary code review framework.

\subsubsection{Alignment Analysis}
\label{subsubsec:alignment-analysis}
PR-issue alignment evaluates whether a PR accurately implements the requirements specified in the associated issue~\cite{Isik2025}. Isik et al.~\cite{Isik2025} formalized this concept by defining four distinct alignment categories: ``exact'' where a PR fully addresses requirements, ``tangling'' where a PR includes unrelated changes, ``missing'' where the PR fails to fully address the issue, and ``missing and tangling'' combining both deviations. Missing implementations serve as indicative signs of technical debt~\cite{Tom2013, Alves2016}, while tangling PRs hinder review effectiveness and defect detection by creating noise that obscures critical changes~\cite{Herzig2016, Tao2015}. The theoretical origins of this alignment derive from literature on tangling commits. PR-issue alignment aims to help PR reviewers identify scope deviations early and ensure their modifications align with the intended requirements. Historically Herzig and Zeller~\cite{Herzig2013} first defined a tangling commit as combining unrelated changes, demonstrating that up to 20\% of bug fixing commits were tangled~\cite{Herzig2016}. Subsequent research focused on preventing and untangling such commits. Kirinuki et al.~\cite{Kirinuki2014} proposed a template based IDE mechanism to mitigate tangling. To untangle existing commits, Dias et al.~\cite{Dias2015} developed \textit{EpiceaUntangler} to automatically cluster tangled changes, and Yamashita et al.~\cite{Yamashita2020} proposed \textit{ChangeBeadsThreader} to visualize fine grained changes for manual clustering. Later, Li et al.~\cite{Li2022UTANGO} introduced \textit{Utango}, utilizing hierarchical agglomerative clustering to produce code change embeddings. Because prior work focused on binary classification at the commit level, Isik et al.~\cite{Isik2025} extended this paradigm to the broader PR-issue relationship. Their taxonomy captures the full spectrum of alignment, demonstrating through manual labeling and zero-shot LLM prompting the substantial potential of integrating LLMs into automated alignment mechanisms.

Building upon these findings, the \textit{Alignment Analysis Agent} automates PR-issue alignment analysis. The agent begins by retrieving the detailed context of the issue, including the title, description, and specific acceptance criteria, alongside the corresponding PR details and source code differences using registered tools. Using this context, the agent classifies the PR into one of the four established PR-issue alignment categories. To provide deeper understanding, the agent highlights the specific tangling lines and irrelevant additions directly within the interactive \textit{diff-map}. Following this analysis, the agent adds a report to the PR review panel. This report contains the line numbers of tangling modifications and provides explicit details regarding any missing implementations. By exposing these discrepancies clearly, the system enables PR reviewers to demand necessary revisions.

\subsubsection{Bug Proneness Analysis}
\label{subsubsec:bug-proneness-analysis}
Bug proneness analysis evaluates the likelihood that PR-diff modifications will introduce defects into the existing system. Although defect identification represents a primary objective of code review~\cite{Bacchelli2013}, effectively achieving this requires a deep understanding of the codebase. Subtle logical errors and edge cases are often indistinguishable during manual review, and studies highlight the frequent fallibility of manual defect identification tasks. Kononenko et al.~\cite{Kononenko2015} revealed that 54\% of code reviews in the Mozilla project failed to identify bugs present in approved commits. Similarly, Czerwonka et al.~\cite{Czerwonka2015} found within Microsoft that a mere 15\% of review comments specifically pointed to potential defects. The strategy of analyzing historical signals to estimate change reliability addresses this critical gap and is well founded in software engineering research. Early studies established the predictive power of version history and code metrics. For instance, Nagappan and Ball~\cite{nagappan2005use} demonstrated that relative code churn is a strong indicator of defect density, while Kim et al.~\cite{kim2007predicting} utilized cached history to identify fault prone hotspots for prioritized verification. Building on these foundations, recent work shifted toward Just-In-Time (JIT) defect prediction, where models analyze specific code changes rather than entire files. Hoang et al. introduced frameworks such as DeepJIT~\cite{hoang2019deepjit} and CC2Vec~\cite{hoang2020cc2vec}, which apply DL to represent code diffs and commit messages to automate risk estimation at the change level. To ensure this automation supports developers, research expanded to explainability and granularity. Pornprasit and Tantithamthavorn~\cite{pornprasit2021jitline} developed JITLine to provide finer grained localization of defects, and Khanan et al.~\cite{khanan2020jitbot} integrated these insights into developer workflows via JITBot. Most recently, LLMs advanced this domain by generating natural language rationales for risk. Abreu et al.~\cite{abreu2024moving} implemented diff risk scoring with LLMs to manage release deployment at an industrial scale. 

Utilizing these methodological advancements, the \textit{Bug Proneness Analysis Agent} automates the estimation of change reliability within our proposed code review workflow. This agent retrieves essential signals including hotspot files, code churn metrics, historical defect density, dependency sensitivity, and changes to error handling paths using tools. By synthesizing these metrics, the agent calculates an overall risk score and estimates where the new changes are likely to introduce failures~\cite{hoang2019deepjit}. Importantly, the agent reports this risk alongside explicit reasons that the PR reviewer can audit. For example, the agent explicitly flags files exhibiting high churn rates and prior incident histories. The agent also bounds these assertions with uncertainty reporting, indicating low confidence when historical data is sparse. Following this calculation, the \textit{Bug Proneness Analysis Agent} outputs a detailed risk evaluation report into the dedicated panel in the PR review panel.

\subsubsection{Impact Analysis}
\label{subsubsec:cia}
CIA evaluates the broader ripple effects of a PR across the software architecture. During manual inspection, PR reviewers easily spot localized syntactic errors but struggle to accurately predict side effects on distant project modules~\cite{Li2013}. This difficulty stems from complex interdependencies where a single modification can inadvertently trigger severe regression defects in dependent components~\cite{Fowler2004}. While traditional CIA offers a methodological approach to estimate these potential effects~\cite{Sun2010, Bohner1996}, performing this analysis manually requires constructing a complex mental model of the entire execution flow, which overwhelms PR reviewers~\cite{Blischak2016}. Göçmen et al.~\cite{Gmen2025} emphasize that traditional platforms fail to visualize these change impacts. Recent research underscores the necessity of multidimensional impact classifications as systems grow in architectural complexity. Bakhtin et al.~\cite{Bakhtin2025} propose utilizing network centrality metrics within service dependency graphs to assess architectural criticality. Cerny et al.~\cite{Cerny2025} emphasize that CIA must extend to infrastructure-level changes in microservices. Beyond structural propagation, detecting subtle behavioral shifts remains a challenge. Jayasuriya et al.~\cite{Jayasuriya2024} demonstrate that semantic breaking changes in APIs often pass syntactic checks yet cause significant downstream failures. On the operational front, Nejati et al.~\cite{Nejati2024} introduce an impact knowledge graph to trace modifications in build specifications. Furthermore, automated adaptation techniques have proven essential for managing cross module impacts. Nielsen et al.~\cite{Nielsen2021} and Scherzinger et al.~\cite{Scherzinger2021} illustrate the value of semantic patches for evolving libraries and databases, while Haryono et al.~\cite{Haryono2020} show that learning from single examples effectively resolves deprecated API usages.

Using these techniques, the \textit{Impact Analysis Agent} automates CIA within our framework. This agent constructs a CIA report to evaluate which downstream modules the proposed patches affect and to calculate the architectural centrality of the modified components. The agent analyzes API and schema alterations, user observable behavioral shifts, operational performance implications, and backward compatibility concerns. Consequently, the agent correctly classifies a seemingly minor diff that alters a default timeout or serialization format as a high impact modification because it affects numerous system modules. The generated report includes a clear classification of impact types, including behavioral, interface, and operational impacts, alongside the specific scope of each effect ranging from local modules to cross service boundaries. The \textit{Impact Analysis Agent} subsequently integrates this evaluation directly into the dedicated PR review panel.

\subsubsection{Runtime Analysis}
\label{subsubsec:runtime-analysis}

Unlike algorithmic complexity analysis, runtime analysis here refers to the behavioral execution of the submitted code changes: the \textit{Runtime Analysis Agent} runs the code in isolated environments and collects execution logs, application traces, and rendered outputs to produce objective evidence about how the software behaves at execution time.
Runtime analysis evaluates the dynamic execution behavior of PR modifications to extract evidence from execution logs, application traces, and rendered web pages. MacLeod et al.~\cite{MacLeod2018} demonstrate that the manual inspection of code changes is hindered by significant cognitive load and code understanding barriers, which runtime analysis specifically aims to resolve. By actively simulating the development workflow to verify if the implementation functions correctly, runtime analysis eliminates the need for developers to manually simulate complex execution flows in their minds. Furthermore, PR reviewers may debate the subjective runtime behavior of the proposed modifications, necessitating objective verification mechanisms to reach a consensus. Following this line of thought in academia, researchers leverage containerization technologies like \textit{Docker} to execute code in a virtual environment. \textit{Docker} provides an isolated virtual environment where models can safely run code modifications, allowing LLMs to utilize this as an executable tool~\cite{merkel2014docker}. With the advancement of LLMs, \textit{Docker} began to be used as an executable tool for LLMs to run code. Following this approach, Dou et al.~\cite{Dou2025} developed \textit{MPLSandbox}, a multi-programming-language sandbox designed to provide unified compiler feedback and secure execution environments for LLMs. Tufano et al.~\cite{Tufano2024} introduced \textit{AutoDev}, an AI-driven development framework that enables agents to perform complex build and execution operations within a secure repository environment. Huang et al. proposed \textit{TraceCoder}, a trace driven multi agent framework for automated debugging, demonstrating how fine grained runtime traces provide deep insights into internal execution states to facilitate precise error localization~\cite{Huang2026}. Pabba et al. presented \textit{SemAgent}, a semantics aware program repair agent that effectively utilizes execution traces to retrieve relevant context for program repair and bug localization tasks~\cite{Pabba2025}. Rondon et al. evaluated agent based program repair at Google, highlighting that integrating execution feedback within agentic workflow demonstrates practical usability and scalability in industrial settings~\cite{Rondon2025}.
Within the commercial landscape, technology companies have released several software engineering agents that utilize integrated runtime environments to automate development cycles, including \textit{Devin}~\cite{devin}, \textit{Jules}~\cite{jules}, and \textit{Replit}~\cite{replit}. For example, \textit{Jules}, developed by Google, runs each task inside a secure and short lived virtual machine to ensure safety, and \textit{GitHub Copilot cloud agents}~\cite{githubCopilotCloudAgent} follow a similar sandboxed execution strategy.

However, a critical reliability concern for execution-based analysis is test non-determinism: tests that may pass or fail inconsistently across runs against unchanged code, commonly known as flaky tests~\cite{luo2014flaky}.
Luo et al.~\cite{luo2014flaky} identify asynchronous waits, concurrency, and test order dependency as the three dominant root-cause categories across 201 commits that likely fix flaky tests in 51 open-source projects, while Lam et al.~\cite{lam2019rootcausing} show in an industrial setting that even a small number of distinct flaky tests can cause a substantial fraction of CI build failures.
When the \textit{Runtime Analysis Agent} collects execution evidence, a single-run outcome may therefore not reflect the stable behavioral state of the submitted code; the agent must employ multi-run aggregation or explicit flakiness classification before treating execution results as reliable review evidence~\cite{lam2020longitudinal}.

Building upon these academic advancements and industrial applications, our proposed framework employs a designated \textit{Runtime Analysis Agent}. This agent operates to create collective, objective, and reproducible evidence regarding the behavior of the PR. When feasible, the \textit{Runtime Analysis Agent} runs targeted executions within a containerized virtual environment it provisions to collect execution logs, application traces, and interface screenshots that directly compare system behavior before and after the change. This agent can also be accessed through the \textit{PR Review Agent} during the \textit{AI-Assisted Code Review} stage by PR reviewers. Consequently, PR reviewers can ask the agent to perform code executions during the review process. For instance, a PR reviewer may wonder about the performance stability of a recently implemented sorting algorithm under extreme load conditions. In this case, the PR reviewer instructs the \textit{Runtime Analysis Agent} to execute the sorting algorithm against randomized datasets of varying sizes to stress-test its performance characteristics. The agent subsequently captures and returns execution traces, diagnostic logs, and performance metrics alongside the analysis. This empirical evidence allows both the PR reviewer and the PR author to understand architectural tradeoffs objectively, effectively replacing speculation during code review with verifiable evidence grounded in actual program execution results.

\subsubsection{Summary Generation}
\label{subsubsec:summary-generation}
Summary generation synthesizes the disparate findings from the \textit{Alignment Analysis Agent}, the \textit{Bug Proneness Analysis Agent}, the \textit{Impact Analysis Agent}, and the \textit{Runtime Analysis Agent} into a cohesive and structured summary. Utilizing these context grounded techniques, our \textit{Summary Generation Agent} constructs a technical overview by synthesizing alignment categories, risk scores, impact classifications, and behavioral evidence to define the modification rationale, potential architectural regressions, and test results. To prevent hallucinated claims, each analysis agent produces structured findings containing claims paired with explicit evidence references (file paths, line numbers, metric values) and confidence indicators. The \textit{Summary Generation Agent} preserves these claim-evidence pairs in its output, surfacing low-confidence markers visibly to PR reviewers rather than smoothing over uncertainty. When upstream agents produce contradictory findings, e.g., an exact alignment classification alongside significant cross-module impact, the summary surfaces the tension explicitly rather than resolving it silently. By embedding this summary directly into the PR review interface, our framework ensures PR reviewers base their decisions on an explicitly grounded analytical summary.

Once the \textit{PR Augmentation} stage is completed, our code review framework proceeds to the \textit{Reviewer Selection} stage for newly created PRs. In this case, the framework utilizes the analytical results to identify and assign the optimal PR reviewers based on the evaluated technical factors. If the PR is a revision to an existing PR, the workflow bypasses the selection phase and moves directly to the \textit{AI-Assisted Code Review} stage once the assigned PR reviewers receive notifications and commence their reviews.
\subsection{Reviewer Selection}
\label{subsec:reviewer-selection}
Once unassigned PRs complete the \textit{PR Augmentation} stage, our framework proceeds to the \textit{Reviewer Selection} stage to identify optimal PR reviewer candidates as illustrated in Figure~\ref{fig:ai_code_review_framework_workflow}. This stage aims to balance workload and knowledge distribution while identifying PR reviewers with code familiarity and substantial experience for critical modifications. In traditional code review systems, achieving this balance presents several challenges. Assigning reviewers with specific module familiarity and general review experience is important for defect detection and efficient feedback~\cite{Kononenko2015,Rahman2017}. However, suboptimal assignments can lead to high workloads~\cite{Ruangwan2019}, causing rejected invitations~\cite{Hirao2016} or stalled PRs~\cite{Gon2024}. Furthermore, balancing immediate QA with knowledge distribution remains difficult~\cite{Bacchelli2013,Ebert2019,Kovalenko2020}, and manual selection can be compromised when PR authors repeatedly assign familiar peers to bypass rigorous inspection~\cite{Doan2022}. To resolve these challenges, our system provides a \textit{Reviewer Suggestion Tool} to recommend PR reviewers by optimizing technical expertise, historical experience, knowledge sharing, and current workload capacity.

\subsubsection{Reviewer Recommendation}
\label{subsubsec:reviewer-recommendation}
Reviewer recommendation automates the identification of suitable PR reviewers based on technical and socio-technical factors. In traditional code review systems, developers assign PR reviewers manually without adequately considering expertise, experience, business context, and knowledge distribution, which can delay integration and reduce defect detection. To address these manual assignment inefficiencies, researchers focused on various automated reviewer recommendation algorithms. Early algorithms primarily targeted specific PR reviewer expertise. Thongtanunam et al.~\cite{Thongtanunam2015} introduced \textit{REVFINDER}, which leverages file location similarities from previously reviewed file paths to recommend suitable reviewers. Balachandran~\cite{Balachandran2013} developed \textit{Review Bot} to automate code review assignments by generating recommendations based on the change history of source code lines. Xia et al.~\cite{Xia2015} improved accuracy by combining text mining of review comments with file location analyses to suggest appropriate PR reviewers for new changes. Sülün et al.~\cite{Sln2019} utilized artifact traceability graphs to improve reviewer suggestion accuracy. Sülün et al.~\cite{Sln2021} later enhanced this approach by incorporating link recency. Subsequent techniques shifted toward broader reviewer experience. Hannebauer et al.~\cite{Hannebauer2016} empirically compared multiple recommendation algorithms to automatically suggest reviewers based on their historical experience. Rahman et al.~\cite{Rahman2016} proposed \textit{CORRECT}, which identifies reviewers by evaluating their cross project and technology specific experience. Asthana et al.~\cite{Asthana2019} developed \textit{WhoDo} to automate reviewer suggestions at scale by incorporating developer workload into the recommendation process. Al Zubaidi et al.~\cite{alzubaidi2020} formulated workload aware reviewer recommendation as a multi objective search problem to balance expertise and reviewer availability. Mirsaeedi and Rigby~\cite{mirsaaedi2020} proposed a recommendation approach to mitigate developer turnover by balancing technical expertise, workload, and knowledge distribution across the team. Rebai et al.~\cite{Rebai2020} introduced a multi objective approach balancing expertise, reviewer availability, and the history of collaborations to optimize the review process.

While existing recommendation studies present mature algorithms in the literature, our visionary code review framework integrates a dedicated \textit{Reviewer Suggestion Tool} to automate socio-technical PR reviewer selection based on previous studies. First, this \textit{Reviewer Suggestion Tool} gathers the required information including repository details, reviewer profiles, current workloads, and additional contextual data. Then, this tool utilizes configurable recommendation settings to suggest optimal reviewers based on expertise, experience, code familiarity, and workload while ensuring knowledge sharing and preventing bad review practices. While some recent studies utilize LLMs for PR reviewer recommendation, our framework uses existing recommendation techniques because they offer a better balance between computational efficiency and token cost tradeoffs. Developers can interact with this tool using the custom panel integrated directly within the PR review menu. For example, developers can interact with the tool to prioritize expertise and experience for critical modules or prioritize the knowledge sharing aspect for less critical components. Once developers select the appropriate reviewer candidates, they send invitations through the platform. Finally, once the selected reviewers accept the invitation, our AI-powered code review framework continues with the \textit{AI-Assisted Code Review} stage.
\subsection{AI-Assisted Code Review}
\label{subsec:ai-assisted-code-review}

Once the invited, responsible PR reviewers begin reviewing the PR, our framework continues with the \textit{AI-Assisted Code Review} stage as shown in Figure~\ref{fig:ai_code_review_framework_workflow}. In this stage, our system generates a \textit{diff-map}, which organizes PR code changes around logical units such as functions, classes, and modules. Rather than presenting changes as raw sequential file differences, the diff-map assigns each unit a name and reference point that ties the code to its evidence. The system then utilizes the main \textit{PR Review Agent} to enable PR reviewers to review the PR using natural language. Traditional code review is often ineffective, with empirical studies indicating that 54\% of reviews fail to detect bugs due to change understanding barriers and approximately 44.47\% of feedback is non-useful~\cite{Kononenko2015, Rahman2017}. These challenges are further compounded by cognitive load from large diffs~\cite{Czerwonka2015}, time pressure~\cite{MacLeod2018}, and toxic communication patterns that degrade team collaboration~\cite{Sarker2023}. To address these challenges, our AI-assisted code review system proposes specialized LLM agents to provide verifiable evidence, contextual insights, and interactive remediation, aiming to transform code review from a manual memory task into a structured retrieval and dialogue process.

In order to address these challenges and enhance the code review process, our framework orchestrates multiple specialized agents around a unified interaction layer. The \textit{PR Review Agent} functions as the central natural language interface that manages the entire review lifecycle and delegates complex inquiries to subordinate agents. The \textit{PR Review Agent} uses the \textit{PR Augmentation Agent} and its subagents as accessible tools to provide analytical context to the PR reviewer. Working alongside this orchestrator, the \textit{Automated PR Review Agent} rigorously inspects the modifications to identify syntax errors, policy violations, and implementation defects. Simultaneously, the \textit{Fix Suggestion Agent} generates concrete, executable code patches to resolve the identified issues automatically. Furthermore, the \textit{Toxicity Measurement Agent} evaluates the sentiment of PR reviewer feedback to ensure professional communication standards. Additionally, the \textit{Usefulness Measurement Agent} analyzes the actionable relevance of review comments to prevent superficial discussions. All framework agents interact via the \textit{diff-map}, a multidimensional substrate that anchors analytical reports, execution traces, and requirement alignment markers to specific code segments. This anchoring enables PR reviewers to conduct verifiable, conversational reviews instead of manual code tracing.

\subsubsection{PR Explanation}
\label{subsubsec:pr-explanation}

One of the components of the \textit{AI-Assisted Code Review} stage is the \textit{Explanation Agent}, which assists in code comprehension by answering natural language questions with evidence-carrying responses anchored to specific code locations. This agent aims to resolve the challenge of change understanding by providing immediate contextual clarifications, mitigate the cognitive load of large changesets by summarizing localized logic, and alleviate time pressure by eliminating the need for manual reverse engineering. Early academic efforts framed code comprehension as a question-answering problem, with Liu and Wan~\cite{liu2021codeqa} introducing \textit{CodeQA} to measure basic source code comprehension. Li et al.~\cite{li2024infibench} subsequently developed \textit{InfiBench} to assess free-form question-answering capabilities across a variety of programming tasks. Sahu et al.~\cite{sahu2024codequeries} presented \textit{CodeQueries} to demonstrate the necessity of multi-hop semantic queries over code spans. Hu et al.~\cite{hu2024coderepoqa} introduced \textit{CodeRepoQA} to capture the complexities of multi-turn conversations within code repositories. Applying this to code changes, Tian et al.~\cite{tian2022change} modeled patch understanding as a question-answering task by correlating bug descriptions with code changes. Furthermore, Dinella et al. introduced \textit{CRQBench} to derive code reasoning questions directly from authentic code review comments to isolate and evaluate semantic reasoning.

Building upon these findings, our framework presents the \textit{Explanation Agent} to provide verifiable reasoning during the PR review process. The \textit{Explanation Agent} analyzes the PR description, the linked issue requirements, and the source code modifications to construct precise, evidence-backed answers. PR reviewers interact with this agent through the \textit{PR Review Agent} utilizing a natural language chat interface. When a PR reviewer asks a complex architectural question, the \textit{PR Review Agent} routes the query to the \textit{Explanation Agent}, which then retrieves the relevant context and formulates a structured response. This response includes direct pointers to the \textit{diff-map}, allowing PR reviewers to navigate to the exact code anchors supporting the explanation.
Reviewer questions impose different levels of cognitive demand: maintainability-oriented queries are often resolvable from the local diff, whereas functional-defect queries require change-wide comprehension~\cite{Beller2014}.
Accordingly, the \textit{Explanation Agent} calibrates its retrieval depth and response granularity to the inferred query type, rather than treating every question as equivalent.

\subsubsection{Automated PR Review}
\label{subsubsec:automated-pr-review}

Another agent in the \textit{AI-Assisted Code Review} stage is the \textit{Automated PR Review Agent}. This agent performs a review of the PR code diff to autonomously identify syntax errors, policy violations, and logical implementation defects. By surfacing these issues immediately, this agent specifically aims to solve the challenges of change understanding, time pressure, and managing large changesets. Manually detecting subtle defects in massive PRs overwhelms human PR reviewers and frequently leads to overlooked vulnerabilities, but the \textit{Automated PR Review Agent} provides oversight by systematically analyzing every modified line. As detailed in Section~\ref{subsubsec:automated-code-review}, this agent formulates its findings as actionable review comments securely tied to exact code locations. During the review, the PR reviewer interacts with these automated findings via the \textit{PR Review Agent} using natural language commands. The PR reviewer can ask the system to elaborate on a specific policy violation or request alternative remediation strategies directly through the chatbot panel. These review comments are visually anchored to the \textit{diff-map}, ensuring the PR reviewer can seamlessly verify the identified defect within the broader functional context.

\subsubsection{Fix Suggestion}
\label{subsubsec:fix-suggestion}

To facilitate the immediate resolution of surfaced issues, the \textit{Fix Suggestion Agent} automatically generates executable code patches and minimal change snippets designed to resolve the defects identified during the automated review, as well as those surfaced through manual human feedback. The \textit{Fix Suggestion Agent} aims to address time constraints and extended review turnaround times by shifting the remediation burden from manual human coding to automated patch generation. As detailed in Section~\ref{subsubsec:pr-fix-suggestion}, this agent utilizes LLMs to produce structurally sound code repairs that maintain the original implementation intent. PR reviewers interact with this agent through the \textit{PR Review Agent} by requesting natural language modifications to the proposed patches or by prompting the agent to generate a fix based on a newly submitted review comment. Once the PR reviewer validates and approves a generated repair on the \textit{diff-map}, the system forwards the patch to the PR author as a formal suggestion. The PR author can then review the suggested fix and apply it directly to the branch, ensuring that the PR author maintains ultimate control over the PR.

\subsubsection{Toxicity Measurement}
\label{subsubsec:toxicity-measurement}

Toxic review comments contain aggressive, insulting, or profoundly negative language that attacks the PR author rather than constructively critiquing the code. The presence of such toxic comments may cause side effects, including elevated stress levels, developer burnout, and mental health problems among software engineering professionals~\cite{Sarker2023}. Furthermore, these negative interactions actively destroy team cohesion, discourage newcomer participation, and damage the knowledge-sharing culture essential for collaborative software development~\cite{Ahmed2017}. Sarker et al.~\cite{Sarker2023} developed \textit{ToxiCR} to automatically identify and classify toxic communications within code review interactions. Zhuo et al.~\cite{Zhuo2025} investigated combating toxic language by reviewing various LLM-based strategies tailored specifically for software engineering environments. Imran et al.~\cite{Imran2025} studied understanding and predicting derailment in conversations on GitHub to preemptively identify discussions devolving into toxicity. Mishra and Chatterjee~\cite{Mishra2024} explored the utilization of ChatGPT for toxicity detection to evaluate the efficacy of generative models in identifying offensive developer communications. Çağlar et al.~\cite{ccauglar2025automated} leveraged LLMs to identify specific review comment smells including toxic comments. While traditional toxicity detection systems often rely on static keyword lists or ML classifiers, the identification and mitigation of context-dependent emotional language in real-time is advanced by the capabilities of LLMs~\cite{Mishra2024}.

Building on these studies, our framework proposes a \textit{Toxicity Measurement Agent} to actively monitor and regulate the sentiment of PR review comments. This agent is integrated with the \textit{PR Review Agent}, enabling proactive remediation workflows when toxic or aggressive language is detected in review comments.

\subsubsection{Usefulness Measurement}
\label{subsubsec:usefulness-measurement}

PR review comment usefulness defines the degree to which reviewer feedback provides actionable, constructive, and relevant guidance to help the PR author improve the code modifications. When PR reviewers submit non-useful comments focusing on trivial stylistic preferences or vague opinions, they cause process inefficiencies and distract from critical flaws. The accumulation of such superficial comments delays merge times, exacerbates time pressure, and causes PR authors to waste effort on trivial revisions~\cite{Bosu2015}. Rahman et al.~\cite{Rahman2017} predicted the usefulness of code review comments by analyzing textual features and developer experience levels. Yang et al.~\cite{Yang2023} introduced \textit{EvaCRC} to systematically evaluate the quality and actionability of generated code review comments. Ahmed and Eisty~\cite{Ahmed2025} evaluated the usefulness of code review comments by comparing textual feature-based and featureless approaches. Çağlar et al.~\cite{ccauglar2025automated} categorized useful review intents using automated LLM classification. Li et al.~\cite{Li2022AUGER} developed \textit{AUGER} to automatically generate high-quality, useful review comments utilizing pre-training models. The automated assessment of comment utility and actionability became highly effective due to the advanced contextual comprehension provided by LLMs.

Building on these studies, our framework implements a \textit{Usefulness Measurement Agent} to ensure that all review comments, whether generated by human PR reviewers or the \textit{Automated PR Review Agent}, actively contribute to code quality. This approach mirrors the findings of Li et al.~\cite{Li2022AUGER} in \textit{AUGER}, which emphasizes that high-quality automated feedback must be as actionable as expert human critique to be effective. This agent is integrated into the comment box, continuously analyzing the technical depth and actionability of the text while the PR reviewer drafts their response. The agent explicitly identifies superficial remarks, such as simple complaints about variable naming conventions without providing alternatives, to prevent bikeshedding. Depending on organizational configurations, the system can prompt the PR reviewer to elaborate on vague statements or block the posting of entirely useless feedback. Furthermore, the \textit{Usefulness Measurement Agent} works in collaboration with the \textit{PR Review Agent} to assist the reviewer. For instance, a PR reviewer can draft a brief, high-level concern and ask the \textit{PR Review Agent} to make the comment actionable, triggering the system to expand the thought into a detailed, constructive critique with a specific remediation proposal.

\subsubsection{PR Review and Composing Comments}
\label{subsubsec:pr-review-comments}

Once the PR reviewer begins the review, they access a comprehensive suite of analytical evidence directly within the platform. The PR reviewer sees the complete analysis report, which includes alignment analysis, bug proneness, runtime analysis, and impact analysis, displayed neatly on a custom user interface panel adjacent to the code. These insights and analytical remarks are also visually embedded as interactive anchors directly on the \textit{diff-map}. The PR reviewer navigates this extensive data using the \textit{PR Review Agent} through a dedicated panel. Through this chat interface, the PR reviewer can interrogate the findings, ask for clarifications regarding specific risk scores, or request deeper investigations into flagged dependencies. For example, the PR reviewer may open the chatbot, reference a specific function on the \textit{diff-map}, and instruct the \textit{Runtime Analysis Agent} to execute the code under specific edge-case conditions to verify its stability.

The integration of writing comments, toxicity measurement, usefulness evaluation, and fix suggestion capabilities creates a robust and psychologically safe review ecosystem. When PR reviewers compose their feedback, they are continuously supported by real-time toxicity and usefulness checks, ensuring every submitted comment is both respectful and technically valuable. The \textit{Toxicity Measurement Agent} actively prevents interpersonal conflicts by neutralizing aggressive language before it reaches the PR author. Simultaneously, the \textit{Usefulness Measurement Agent} eliminates process waste by demanding actionable clarity, thereby preventing endless debates over trivial stylistic preferences. If a PR reviewer identifies a flaw, they do not need to manually write out the corrected code. Instead, the PR reviewer leverages the \textit{Fix Suggestion Agent} to automatically generate a fix patch. This generated patch is attached directly to the review comment. The PR author receives a professional, polite, and actionable piece of feedback complete with a one-click implementation solution. Consequently, this mechanism aims to reduce the cognitive friction of the review process, accelerate the overall integration timeline, and foster a deeply collaborative engineering culture.

Once the PR reviewers complete their review and submit a decision to either approve or reject the proposed changes, our framework continues to the subsequent workflow stages. If the PR is approved or rejected, the process advances to the \textit{PR Retrospective} stage. Conversely, based on the feedback received, the PR author may choose to revise and develop the PR further. In this scenario, the PR author re-enters the \textit{Implementation} and subsequent development stages to address the PR reviewer feedback. Alternatively, the PR author may elect to abandon the PR, thereby finalizing the PR lifecycle.
\subsection{PR Retrospective}
\label{subsec:pr-retrospective}
Once the PR is approved or rejected following the \textit{AI-Assisted Code Review} stage, our visionary code review framework continues to the \textit{PR Retrospective} stage as illustrated in Figure~\ref{fig:ai_code_review_framework_workflow}. The PR retrospective stage constitutes the mechanism by which our visionary framework achieves continuous improvement within the code review process. This stage aims to summarize the lifecycle of each PR, enabling humans to track modifications and allowing agents to continuously improve their analytical capabilities. During this stage, a \textit{Review Summary Generation Agent} produces a technical summary while a \textit{Review Metric Computation Tool} collects review data about the PR to preserve repository memory and facilitate decision tracking. By recording architectural rationales and identifying procedural bottlenecks, this stage ensures continuous system improvement and provides context for future PR reviews. Ultimately, this systematic data collection allows project maintainers to customize agents for unique organizational workflows and repository standards.

\subsubsection{Review Summary Generation}
\label{subsubsec:review-summary-generation}
Our framework begins this stage by generating a PR review summary that automatically condenses the discussions, code changes, and analytical reports into an easily readable format. Prior studies investigated the performance of LLMs across diverse domains. Sun et al.~\cite{Sun2025} highlight that the emergence of LLMs has led to a great boost in the performance of source code summarization techniques. Similarly, Zhang et al.~\cite{Zhang2024} demonstrate through benchmarking that fine-tuning empowers LLMs to produce high quality summaries that rival human written texts. Building upon the proven text generation and summarization capabilities of these models, our framework utilizes a \textit{Review Summary Generation Agent} to preserve useful repository memory. The agent begins by retrieving all relevant details, including the initial issue, the code diff, and the subsequent PR reviewer comments. Then, the agent generates a PR review summary incorporating explicit traceability links. This summary contains the details of the PR and the linked issue, alongside the evolution of analytical results over various commits, capturing runtime evidence, CIA, bug proneness, and alignment analysis. Furthermore, the summary documents PR reviewer selection decisions and the key events that happened during the review process, including critical review comments, negotiated resolutions, and final merge or reject decisions. Crucially, these summaries support machine retrieval, ensuring that future agents can utilize these historical decisions to inform subsequent PR review cycles. The retention, access controls, decay, and reuse policies that govern such retrieval are framework-level concerns we treat as open challenges and discuss further in Section~\ref{subsubsec:privacy-challenges}.

\subsubsection{Review Metrics Storage}
\label{subsubsec:review-metrics-storage}
After the generation of the PR review summary, our framework continues to compute the relevant metrics necessary to continuously improve the code review workflow and customize the agents for the specific repository. Several studies highlight the importance of tracking and analyzing metrics during code review. Rigby et al.~\cite{Rigby2015} demonstrate a mixed methods approach to mine code review data across multiple commit reviews and PRs to understand PR author collaboration patterns. Hasan et al.~\cite{Hasan2021} utilize a balanced scorecard approach in an industrial setting to identify concrete opportunities for improving code review effectiveness through targeted metric tracking. Izquierdo-Cortazar et al.~\cite{Izquierdo-Cortazar2017} further establish how tracking specific metrics directly evaluates and improves code review performance. Finally, Tan et al.~\cite{Tan2025} emphasize the necessity of reflective memory management, showing that synthesizing historical interactions enhances the long term performance of personalized LLM agents. Building on these studies, our framework proposes a \textit{Review Metric Computation Tool} to extract actionable data from the completed PR review cycle. This tool computes essential velocity and quality metrics such as review latency, comment usefulness, and the number of review iterations required for approval. By systematically storing this quantitative data alongside the qualitative review summaries, the system empowers engineering managers to identify workflow inefficiencies. Furthermore, instantiating organizations may use these metrics and the qualitative review summaries to customize specialized agents to their quality standards and operational expectations. The mechanics of any such feedback loop are framework-level safeguards. They include human approval, the scope of agent-behavior modification, privacy controls over reviewer and codebase data, and right-to-erasure handling. We surface these as open challenges and discuss them in Sections~\ref{subsubsec:accountability-issue} and~\ref{subsubsec:privacy-challenges}.

\section{Discussion}
\label{sec:discussion}

This section evaluates the operational limitations, risks, and potential implications for practitioners, as well as future research directions for researchers, about the proposed visionary AI-powered code review framework. Section~\ref{subsec:future-challenges} discusses implementation challenges, ranging from technical failures such as model hallucination and cascading error propagation to privacy concerns, unpredictable economic costs, and negative side effects including professional accountability dilemmas and automation bias. Section~\ref{subsec:implication-for-practitioners} details the implications for software practitioners, emphasizing the changing roles of stakeholders, modular adoption strategies, and the shift toward internal platform development for code review. Section~\ref{subsec:implications-researchers} explores future research directions, highlighting the requirement to redefine review metrics, optimize the user experience of AI-assisted review, investigate context enrichment mechanisms, and analyze the long-term consequences of AI-powered code review.

\subsection{Challenges, Risks and Limitations}
\label{subsec:future-challenges}
While the proposed AI-powered code review framework aims to address modern code review bottlenecks, transitioning to a multi-agent system introduces profound socio-technical risks. The stochastic nature of underlying models causes hallucinations and context degradation. Additionally, their inability to generalize across proprietary repositories threatens localized accuracy. Within interconnected pipelines, isolated inaccuracies can transform into cascading errors that compound systemic failures and unpredictable economic costs. Furthermore, the inherent opacity of these models creates severe deficits in transparency, accountability, and privacy. These deficits complicate governance and data security. Finally, excessive automation threatens to degrade HITL oversight through automation bias, deteriorate team knowledge sharing, and complicate system evaluation. Consequently, this section evaluates these limitations by detailing their impacts and exploring targeted mitigation strategies.

\subsubsection{Bias and Inaccuracy in LLM Predictions}
As Chen et al.~\cite{Chen2021} demonstrate, while the integration of models specifically fine-tuned on source code improves baseline syntactic generation, the inherent stochasticity of LLMs introduces hallucinations that act as critical blockers within the proposed \textit{AI-Assisted Code Review} stage. Rather than merely producing formatted text, these models function fundamentally as stochastic parrots that stitch together probabilistic patterns without actual semantic grounding, a limitation detailed by Bender et al.~\cite{Bender2021}. In practical code generation scenarios, Zhang et al.~\cite{zhang2025llm} and Chen et al.~\cite{Chen2024} categorize how this stochasticity manifests as specific, non-syntactic defects, such as the invocation of non-existent APIs or the fabrication of phantom variables. Within our multi-agent architecture, such hallucinations severely disrupt the causal chain of the review workflow. For example, if the \textit{PR Review Agent} probabilistically fabricates a false positive security vulnerability, this hallucinated risk directly triggers the \textit{Fix Suggestion Agent} to generate an unnecessary, invalid patch. Consequently, instead of accelerating integration, the system forces the human reviewer to expend significant cognitive effort untangling AI-generated noise. This compounding failure demonstrates that model hallucination is not a secondary inconvenience but a primary structural threat that can systematically corrupt automated remediation efforts if explicit human oversight is bypassed.

Beyond hallucinations, context degradation represents a fundamental constraint that paralyzes the architectural analysis of large, real-world PRs during the PR Augmentation stage. Although contemporary LLMs boast expanding maximum token limits, their ability to robustly retrieve and utilize information degrades significantly across long input contexts. Liu et al.~\cite{Liu2024} empirically demonstrated this ``Lost in the Middle'' phenomenon.
They revealed a U-shaped performance curve in which models fail to access relevant information located in the center of their context window. This degradation directly threatens the operational viability of the Impact Analysis Agent. When tasked with analyzing a massive 50-file refactoring PR, the agent requires complete retention of initial module definitions to accurately trace downstream dependencies. Because of context degradation, the \textit{Impact Analysis Agent} may lose track of critical interface changes provided early in its prompt, causing it to fail to detect severe cross-module side effects. Consequently, the agent might incorrectly categorize a breaking architectural change as safe, providing a dangerously flawed analytical foundation for the subsequent reviewer selection and review stages. Therefore, token constraints dictate that our framework must incorporate explicit token threshold warnings and mandate manual human mitigation for changes exceeding the model's reliable retrieval capacity.

Finally, the intrinsic opacity and training biases of LLMs systematically undermine the verifiability and security of the automated workflow. The ``black box'' nature of neural architectures creates a severe interpretability deficit, meaning the exact inferential steps used to generate a review comment remain hidden from the end user, as defined by Liu et al.~\cite{Liu2024}. This lack of transparency has profound practical implications; Davila et al.~\cite{Davila2024} emphasize that software practitioners cannot establish trust in AI-driven reviews without transparent, verifiable reasoning. Because reviewers cannot mathematically verify the logic behind automated decisions, the \textit{Automated PR Review Agent} might erroneously flag a secure, proprietary cryptographic implementation simply because its syntax diverges from the open-source patterns overrepresented in the model's training data. Furthermore, large-scale security evaluations by Pearce et al.~\cite{PearceHammond2025} and Tihanyi et al.~\cite{Tihanyi2024} reveal that AI-generated code consistently reproduces specific vulnerabilities, finding that between 40\% and 51.24\% of generated programs contain security flaws frequently present in uncurated training corpora. Within our framework, this bias means that the \textit{Fix Suggestion Agent} could autonomously inject known security vulnerabilities into the repository while attempting to resolve a trivial defect. This inherent security risk solidifies the absolute necessity of our HITL quality gates, ensuring that no AI-generated code modification or risk assessment bypasses accountable human validation.

\subsubsection{Limited Generalization Across Different Software Projects}
Generalization failure across diverse software projects comprises an architectural vulnerability. This failure originates from severe distribution shifts between the public open-source repositories used for training and the highly idiosyncratic nature of proprietary software. As Zhang et al.~\cite{Zhang2016} demonstrate, large neural networks used in LLMs often achieve high accuracy by memorizing their training data rather than learning generalizable abstractions. Furthermore, Koh et al.~\cite{WeiKoh2021} establish that in-the-wild distribution shifts substantially degrade the out-of-distribution accuracy of ML systems. Within the \textit{AI-Assisted Code Review} stage, this memorization bottleneck may compromise the \textit{PR Review Agent}. When an agent trained predominantly on standard open-source patterns encounters a proprietary frontend architecture requiring complex local state configurations and custom build scripts, the underlying distribution assumption breaks. Consequently, the agent may fail to comprehend the custom architecture and erroneously flag valid proprietary patterns as anti-patterns.

The inability to generalize across diverse programming languages and architectural paradigms may also harm the reviewer selection process. Ray et al.~\cite{Ray2014} demonstrated, through a comprehensive analysis of 729 projects comprising 80 million lines of code, that programming language design inherently alters code quality characteristics and defect proneness. If the \textit{Reviewer Selection Tool} applies generic heuristic weights learned from statically typed Java repositories to evaluate a dynamically typed Python microservice, it may suggest a suboptimal reviewer for the PR. This causes the system to misroute the PR to an underqualified reviewer, compromising the overall review quality. To prevent these localized failures, we propose two primary mitigation strategies. First, our framework cannot rely strictly on zero-shot inference; the architecture should integrate transfer learning, a technique surveyed by Pan and Yang~\cite{Pan2010} that adapts models across differing feature spaces and data distributions. Second, organizations should establish active feedback loops by leveraging the \textit{PR Retrospective stage}. Because continuous model fine-tuning often incurs high computational costs, the framework utilizes human corrections to iteratively update project-specific configuration files, such as Agents.MD, which define localized architectural skills and constraints. Injecting these explicit, repository-specific rules directly into the agent context window provides a computationally inexpensive alternative to fine-tuning, ensuring continuous adaptation to proprietary configurations.

\subsubsection{Accumulated Error in Multi-Agent Systems} Accumulated error in multi-agent workflows presents a notable challenge that requires careful architectural consideration to ensure the reliability of the code review framework. In our proposed vision, the review process progresses through sequential stages, where the outputs of upstream agents often inform downstream tasks. Consequently, initial inaccuracies such as factual hallucinations or misclassifications can propagate across the pipeline if left unchecked. Asadi et al.~\cite{asadi2019combating} demonstrate that sequential models can experience compounding errors when early imperfections shift the input distribution for subsequent steps. Within our framework, this could occur if an error originates early in the process and bypasses human validation. For example, if the \textit{PR Detail Generation} agent misinterprets complex authorization update logic as a routine business logic adjustment, it will generate an inaccurate PR description. This flawed context is then passed to the Reviewer Selection agent, which subsequently may assign a backend expert rather than a security expert to the review. Because the assigned reviewer may lack the specific domain expertise required, critical security vulnerabilities might be overlooked during the \textit{AI-Assisted Code Review} stage. While Jimenez et al.~\cite{jimenez2024swebench} note that standalone LLMs currently struggle with complex, multi-file software engineering tasks, our multi-agent orchestration must be carefully designed to prevent such localized misunderstandings from compounding into systemic review failures.

Fortunately, our proposed framework is explicitly designed to mitigate these risks by transitioning away from fully autonomous, black-box pipelines. By retaining humans at key decision points, the architecture provides natural firewalls against error accumulation. Wu et al.~\cite{wu2024autogen} highlight that multi-agent frameworks benefit significantly from explicit grounding and validation mechanisms. To this end, our code review pipeline integrates intermediate validation layers and cross-agent feedback loops. For instance, the PR author can verify and correct the AI-generated PR description before creating a PR, effectively breaking the error chain early on. Through this form of human-AI collaboration, the framework ensures that early-stage inaccuracies are caught and corrected, maintaining the overall integrity and trustworthiness of the automated review ecosystem.

\subsubsection{Evaluating the AI Agents}

Evaluating our framework's agents requires multi-step, interaction-based paradigms rather than isolated per-agent assessments.
When the \textit{PR Review Agent} and \textit{Fix Suggestion Agent} are evaluated independently, critical dependency chains remain invisible.
Standard code generation benchmarks such as HumanEval~\cite{Chen2021} measure isolated text generation and cannot detect how an error in one agent propagates to the next.
For example, if the \textit{PR Review Agent} hallucinates a concurrency vulnerability, the \textit{Fix Suggestion Agent} synthesizes unnecessary lock mechanisms, amplifying the initial error into a cascading downstream failure.
Ji et al.~\cite{Ji2023} confirm that such hallucinations are a pervasive limitation of generative LLM pipelines.
This multi-agent error accumulation poses a greater evaluation challenge than the context-window degradation that Liu et al.~\cite{Liu2024} document, because the downstream agent generates a confident but structurally incorrect result rather than a visibly truncated one.
Karakaya et al.~\cite{karakaya2026understanding} demonstrate that automated evaluation alone cannot capture these dynamics: across 2,604 bot-generated PR comments from an industrial deployment, G-Eval and LLM-as-a-Judge strategies achieve agreement ratios of only 0.44 to 0.62 against developer-provided labels, with near-zero Matthews correlation coefficients after class-imbalance adjustment, because developer labeling reflects organizational priorities rather than purely technical assessments of comment usefulness.

Addressing these challenges requires moving beyond surface-level metrics to multi-dimensional evaluation frameworks.
Torun et al.~\cite{torun2026evaluation} identify five structural obstacles specific to evaluating LLM-based SE tools: absent stable ground truth, multi-dimensional and subjective quality, instability from non-determinism, biases in automated judging, and fragmented evaluation practices.
These obstacles explain why reference-based metrics such as BERTScore by Zhang et al.~\cite{Zhang2019} and BLEURT by Sellam et al.~\cite{Sellam2020} are insufficient proxies for code review agent quality.
Both reduce multi-dimensional review quality to surface-level textual similarity and omit dimensions such as defect detection rate, reviewer calibration, and task convergence.
Evaluation designs must also account for economic feasibility.
Kapoor et al.~\cite{Kapoor2024} caution that optimizing for accuracy alone produces needlessly expensive systems.
Dynamic frameworks such as AgentBench by Liu et al.~\cite{Liu2023} provide an alternative, tracking cost-performance trade-offs systematically across multi-agent evaluation settings.

\subsubsection{Transparency of The System}
The inherent architectural opacity of LLMs acts as a fundamental constraint on reviewer trust, necessitating explicit internal reasoning and tool execution traces within our framework. As defined by Lipton~\cite{lipton2018mythos}, deep neural networks lack internal decomposability, meaning human PR reviewers cannot manually audit the inferential steps that lead to a specific automated decision. When an agent outputs an assessment, the absence of trace-level visibility prevents the reviewer from verifying whether the output stems from genuine architectural analysis or spurious pattern matching. For instance, if the Automated \textit{PR Review Agent} flags a newly introduced asynchronous task queue as a critical race condition but suppresses its intermediate reasoning, the human reviewer cannot distinguish a model hallucination from a legitimate concurrency defect without understanding the actual code change. This issue may force PR reviewers to duplicate the analysis effort, negating the efficiency gains of automated workflows. To resolve this, future implementations of the framework could focus on explicitly rendering the agent's evidence traces, tool outputs, and auditable rationales during the review. Furthermore, Kadavath et al.~\cite{Kadavath2022} demonstrated that LLMs possess an inherent capability for self-calibration to estimate the probability of their own correctness. By outputting these calibrated confidence probabilities alongside the execution traces, the framework can transform un-auditable outputs into more transparent and verifiable artifacts, thereby establishing a more transparent system.

\subsubsection{Accountability Issue}\label{subsubsec:accountability-issue}
Accountability in AI-driven code review is a governance challenge, particularly when these pipelines are deployed in environments where software defects can cause substantial harm. This concern is reminiscent of historical software engineering failures like the Therac-25 incident. In our proposed architecture, the boundary of responsibility can be mitigated with HITL checkpoints. For example, if the \textit{PR Review Agent} generates a confident but fundamentally misleading security report, an over-reliant human developer might approve this assessment during the \textit{AI-Assisted Code Review} stage without thorough verification. In such cases, the locus of liability for a subsequent production breach becomes difficult to determine. Floridi et al.~\cite{Floridi2018} describe this as the ``problem of many hands'' in distributed AI governance. This challenge is further complicated by the fact that traditional fault-based product liability regimes struggle to assign blame to autonomous models~\cite{Buiten2024}. To prevent our framework from operating as an unaccountable black box, the architecture is designed to separate AI generation from human decision making. While agents automate some of the tasks, the final merge authority remains exclusively with human PR reviewers. To mitigate these liability risks in practice, implementations of our vision should enforce system-level guardrails. These guardrails include immutable audit logs, decision traces~\cite{Doshi-Velez2017}, and mandatory approval checkpoints. These mechanisms ensure that every AI-generated classification is permanently tethered to the human operator who authorized its deployment.

\subsubsection{Privacy Challenges of LLM Powered Agents}
\label{subsubsec:privacy-challenges}

Privacy challenges in proposed code review systems represent a barrier to enterprise adoption, specifically regarding the handling of proprietary source code and personally identifiable information. In our proposed architecture, agents such as the \textit{PR Review Agent} and the \textit{Fix Suggestion Agent} require access to codebase context, organizational documents, and communication channels which increases the risk of proprietary data leakage during inference or through logging-related vulnerabilities. He et al.~\cite{he2024emerged} identify that LLM agents are particularly susceptible to prompt injection attacks; for instance, a malicious actor could embed instructions in a PR's code comments to manipulate the \textit{PR Review Agent} into exfiltrating sensitive environment variables or internal logic. Furthermore, the risk of training data exposure remains a technical concern for systems utilizing third-party models. Carlini et al.~\cite{274574} demonstrate that large-scale LLMs can be prompted to reveal verbatim fragments of their training sets, which could include sensitive code snippets if the model provider utilizes inference logs for further training. Beyond technical leakage, the transmission of code to external APIs complicates compliance with stringent legal frameworks such as the General Data Protection Regulation (GDPR) and the California Consumer Privacy Act (CCPA), which mandate strict data residency and right-to-erasure protocols~\cite{gdpr2018general}. To mitigate these risks, the proposed system should implement architectural safeguards that balance technical utility with regulatory requirements. By incorporating privacy-preserving configurations and data-handling protocols, the framework should ensure that organizational security remains intact while leveraging the full potential of agents

\subsubsection{Automation Bias}
Automation bias may compromise the HITL verification mechanism in our proposed code review framework, undermining the verification mechanism designed into the multi-agent orchestration. For instance, when agents like the \textit{PR Detail Generation Agent} and the \textit{Issue Linking Agent} auto-generate detailed, seemingly authoritative PR descriptions and issue links, PR authors are prone to accept these artifacts without rigorous verification and validation. If the \textit{Issue Linking Agent} incorrectly links a minor UI update to an unrelated issue, a PR author experiencing automation bias might blindly approve the PR metadata. Because our framework relies on interconnected stages, this unverified issue link triggers a cascading failure: the PR Augmentation stage will generate flawed summaries and analysis based on the wrong issue context, which subsequently misleads the Reviewer Selection agent into assigning PR reviewers with incorrect domain expertise. This passive dependency exposes a fundamental flaw in assuming HITL mechanisms are inherently robust on their own. Supporting this, Sabouri et al.~\cite{Sabouri2025} observed that while developers may easily accept initial AI outputs, they ultimately retain only 52\% of AI-generated suggestions, highlighting a severe discrepancy between passive initial acceptance and actual long-term code quality. Furthermore, Wang et al.~\cite{Wang2024-2} emphasize that developer trust in AI tools is highly situational, necessitating explicit quality indicators rather than assuming blind reliance.

\subsubsection{Knowledge Sharing Deterioration}
Software development is a collective and knowledge-intensive activity, making the protection and transfer of knowledge essential for maintaining project continuity~\cite{bfsig}. Bacchelli and Bird~\cite{Bacchelli2013} established that while finding defects is the primary motivation for code review, knowledge transfer and team awareness are equally critical outcomes. However, the extensive integration of AI agents within our proposed \textit{AI-Assisted Code Review} and \textit{PR Augmentation} stages introduces a systemic risk of deteriorating these educational benefits. When the \textit{PR Review Agent} provides pre-computed risk reports and the interactive chatbot answers architectural questions directly, PR reviewers may be incentivized to bypass manual code-level inspection. This over-reliance creates an environment conducive to vibe coding, a phenomenon where Fawzy et al.~\cite{Fawzy2025} observed developers relying on AI through intuition and trial-and-error without actually comprehending the underlying implementation. For example, if a junior developer relies entirely on the chatbot to explain a complex PR rather than dissecting the code and debating architectural choices with the PR author, the team's long-term technical capability degrades. Furthermore, this automation directly threatens ``implicit mentoring''. This practice refers to the unstructured guidance senior developers provide during human-to-human PR discussions. Feng et al.~\cite{feng2022implicit,feng2022case} demonstrated the critical nature of this process, finding that 27.41\% of PRs in open-source projects contain embedded educational guidance. By offloading discussion to an AI during the review process, the framework risks damaging this mentorship channel.

To prevent the \textit{AI-Assisted Code Review} stage from becoming an educational bottleneck, potential mitigation strategies should focus on utilizing \textit{Retrospective Analysis} and HITL components to actively promote knowledge sharing. Instead of allowing PR reviewers to simply rubber-stamp the \textit{PR Review Agent}'s summary, future workflows could introduce mechanisms that explicitly prompt human-to-human interaction. For example, a new mechanism could mandate a ``PR Critique'' checkpoint where PR reviewers must contribute architectural insights or mentorship notes directly to the PR author before the merge is unblocked. Such a strategy would help ensure that the efficiency gained during the \textit{PR Augmentation} stage does not silently displace the essential human-centric mentoring that sustains a development team's expertise.

\subsubsection{Economic Impacts and Hidden Costs}
Beyond technical and cognitive challenges, the economic viability of this framework requires careful consideration due to the potential for hidden, non-linear cost accumulations. While individual LLM inferences may appear inexpensive, our proposed architecture relies on sequential multi-agent orchestration, which, without proper safeguards, could transform isolated agent errors into compounding financial overhead. The mechanism for this lies in the framework's interconnected context window: if an agent operating early in the workflow, such as during the \textit{PR Creation} or \textit{PR Augmentation} stage, hallucinates or misclassifies an issue, the orchestration layer might unknowingly pass this flawed premise downstream. Because agents incur costs per token processed and generated, the system risks paying not just for the initial output, but for subsequent agents' attempts to reason over, validate, or remediate that fabricated problem, ultimately resulting in unnecessary computational expenditure.

To illustrate this potential causal chain, consider a scenario during the \textit{PR Augmentation} stage where the \textit{Bug Proneness Analysis Agent} incorrectly flags a standard variable renaming as a critical concurrency risk. This upstream misclassification could trigger an inefficient cascade: the \textit{Runtime Analysis Agent} might consume additional compute resources attempting to simulate the non-existent race condition in a sandbox, and the \textit{Fix Suggestion Agent} subsequently uses tokens generating complex, asynchronous locks. Finally, when the human reviewer during the \textit{AI-Assisted Code Review} stage recognizes the error and rejects the effort, the organization has incurred the compounded token and compute costs of three distinct agents for that specific review thread. Consequently, traditional cost-per-token metrics may not fully capture this multiplier effect, potentially underestimating the broader financial implications of cascaded agent pipelines.

To mitigate this financial unpredictability, organizations should evaluate the deployment trade-offs of their underlying models. Investigating these dynamics, Aryan et al.~\cite{aryan2023costly} demonstrate that the cost-optimal deployment of LLMs involves important trade-offs: organizations typically choose between utilizing vendor-based APIs (which minimize upfront capital but introduce variable API fees and prompt-drift expenses) and building in-house models (which offer predictable inference costs but require substantial initial hardware investments). Furthermore, Aryan et al. highlight that computational costs increase quadratically with context window size, meaning extensive, cascading prompts can become expensive. Therefore, to ensure economic efficiency and prevent unnecessary costs from cascading errors, potential mitigation strategies should explore integrating circuit breakers. Such a mechanism would dictate that if an agent produces an output with a confidence score below a pre-determined threshold, the framework could halt the pipeline, preemptively preventing downstream agents from expending resources on statistically uncertain inputs.
\subsection{Implications for Practitioners}
\label{subsec:implication-for-practitioners}
The transition to an AI-powered code review ecosystem calls on software engineering practitioners to rethink their operational workflows and toolchain strategies. The identified implications may directly impact how organizations adopt this AI-powered code review workflow. Rather than prescribing a rigid model, this section outlines considerations for practitioners to safely integrate the framework based on their context and risk tolerance. Moreover, we discuss redefining stakeholder responsibilities, internal code review platform development, incremental adoption strategies, and dynamic, risk-based review routing.

\subsubsection{Redefining Roles and Responsibilities of Stakeholders}
\label{subsubsec:redefining-roles-responsibilities-stakeholders}

The proposed AI-powered framework does not eliminate human stakeholders but redistributes and transforms their responsibilities across the software development lifecycle. By automating context generation and analysis, the framework shifts the human workload from manual execution to interactive operation and supervisory validation. A detailed discussion of these evolving responsibilities is provided, and Table~\ref{tab:stakeholder-roles} summarizes these changes.

\begin{table}[t]
\centering
\small
\renewcommand{\arraystretch}{1.15}
\caption{Comparison of Stakeholder Roles in Traditional vs. Proposed AI-Powered Code Review Framework}
\label{tab:stakeholder-roles}
\begin{tabularx}{\linewidth}{
>{\RaggedRight\arraybackslash\bfseries}p{0.15\linewidth}
>{\RaggedRight\arraybackslash}X
>{\RaggedRight\arraybackslash}X
}
\toprule
Role & \textbf{Traditional Systems} & \textbf{Proposed AI-Powered System} \\
\midrule
PR Author & Manually writes PR descriptions, establishes issue links, and resolves trivial syntax errors prior to PR creation. & Acts as an interactive operator during the PR Creation stage, using natural language to direct agents in refining descriptions, linking issues, and generating fixes. \\
\midrule
PR Reviewer & Manually reads sequential code diffs to hunt for defects, verify requirements, and author remediation code. & Operates specialized agents via natural language to interrogate architectural context, command runtime executions, and coordinate automated patch generation. \\
\midrule
Project Manager / Team Lead & Manually assigns issues, selects PR reviewers based on intuition, and tracks resolution progress. & Oversees high-level issue management while supervising AI-assisted review outputs and project-wide technical goals. \\
\bottomrule
\end{tabularx}
\end{table}

\textbf{PR Authors:} Traditionally, PR authors manually prepare review context, but our framework shifts them into active operators of AI-generated artifacts. Through natural language, PR authors direct the \textit{PR Creation Agent} to iteratively refine descriptions, correct traceability links, and request alternative code patches. For example, rather than passively verifying a description, an author actively commands the system to search for and swap an incorrect backend issue link with the correct frontend requirement.

\textbf{PR Reviewers:} PR reviewers transition from line-by-line manual defect hunting to operating specialized agents that evaluate analysis and coordinate remediation. Utilizing the \textit{PR Review Agent}, PR reviewers actively interrogate the \textit{Explanation Agent} for architectural context and can instruct the \textit{Runtime Analysis Agent} to execute code under specific conditions. For example, instead of manually authoring a fix for a flagged vulnerability, a PR reviewer commands the \textit{Fix Suggestion Agent} to generate and apply a targeted remediation patch.

\textbf{Project Managers / Team Leads:} Their core responsibilities in issue management and developer assignment remain largely consistent, though their role shifts toward high-level governance of the automated workflow.

\subsubsection{Shift to Internal Platform Development} Practitioners may stop acquiring rigid review tools, which inherently optimize for generic workflows, and instead adopt an internal code review platform paradigm. Rather than deploying monolithic, standalone review bots, organizations should architect the Agent-Orchestrated Collaborative Review framework as a suite of reusable services, modular LLM APIs, centralized observability metrics, and access controls. Such an internal platform approach allows engineering teams to move beyond abstraction and explicitly encode repository-specific policies, risk thresholds, and review conventions directly into the system's architecture. For example, during the framework's \textit{PR Creation} stage, an organization can configure the multi-agent orchestration layer to utilize the \textit{Issue Linking Agent} to enforce localized traceability rules, systematically flagging non-compliant code changes before they reach PR reviewers. Furthermore, this level of internal control is a critical requirement for addressing the privacy, transparency, and accountability risks established in earlier sections; as Wagman et al.~\cite{Wagman2025} emphasize in their evaluation of AI-assisted development, organizations increasingly enforce explicit disclosure requirements and strict review protocols to manage the integration of AI-generated code safely. By building these accountability guardrails directly into the platform's infrastructure, practitioners can enforce compliance systematically. However, while this architectural flexibility transforms the review process into a competitive advantage, it simultaneously introduces a governance and maintenance burden. The ongoing economic costs associated with tuning the agents, updating custom agents, and monitoring system-wide privacy thresholds represent a substantial operational overhead, which may force practitioners to critically evaluate whether the strategic benefits of a heavily tailored review platform outweigh the continuous engineering investment required to sustain it.

\subsubsection{Modular and Incremental Adoption Strategy} Implementation of this framework should avoid a simultaneous, all-encompassing deployment strategy. Instead, practitioners should adopt a modular integration path distinguishing between isolated technical component adoption and broader workflow transformation. By anchoring deployment to specific architectural stages, organizations can selectively enable agents that align with their operational maturity and mitigate challenges like accumulated error in multi-agent orchestration. For instance, deploying the \textit{Risk Analysis Agent} during the \textit{PR Augmentation} stage serves as a component adoption that categorizes changes without altering the human-centric review workflow. Such flexibility supports different kinds of organizational adoption profiles. For instance, a defense contractor constrained by strict regulatory compliance, where error propagation is a critical blocker, may restrict automation to risk analysis to maintain mandatory human approval chains. Conversely, a startup driven by throughput pressure might heavily rely on the agent's preliminary reviews, treating minor hallucinations as secondary concerns compared to release velocity. Furthermore, teams can independently deploy downstream agents while explicitly opting out of earlier stages, such as the \textit{PR Creation} stage, to preserve established IDE configurations. This modular adoptability allows organizations to validate system behavior and assess risk thresholds incrementally without disrupting critical software development lifecycles.

\subsubsection{Dynamic Risk-Based Review Routing} 
Treating all PRs with equivalent rigor may be operationally inefficient; practitioners could instead explore dynamic review protocols as a future extension of the framework's \textit{PR Augmentation} stage. For instance, future implementations could introduce a \textit{Risk Analysis Agent} designed to compute specific risk scores evaluating defect probability and measuring architectural impact. This potential dual-metric system could guide PR review routing decisions, requiring expanded security testing and explicit sign-off by designated code owners. ``Fast Lanes'' could be established for low-risk modifications, though they should ideally still pass automated verification, CI checks, and compliance audits before approval. Conversely, core component changes could trigger ``High-Risk Lanes''. For example, a PR updating localized documentation could route through a fast lane with automated validation. Meanwhile, a PR refactoring an authentication database schema is typically routed to a more rigorous PR review flow.

While dynamic routing may potentially optimize throughput, false risk classification emerges as a potentially critical drawback. If a future \textit{Risk Analysis Agent} misclassifies a high-risk change as low-risk, the system might bypass critical rigorous review processes. This structural danger often outweighs secondary concerns like API token limits. To mitigate this, practitioners are encouraged to avoid treating routing logic as a black box. An additional stage for reviewing routing decisions could catch misclassified PRs. Since AI is not a substitute for human accountability, practitioners exploring these extensions might establish continuous monitoring of PR routing classifications. This oversight ensures dynamic routing accelerates velocity without compromising security or operational integrity.

\subsection{Implications for Researchers}
\label{subsec:implications-researchers}

Beyond the practical considerations for adoption, the proposed framework surfaces several open research problems that require sustained investigation by the software engineering research community. The following subsections identify priority research directions, ranging from workflow heterogeneity and error containment to evaluation methodology, model enhancement, and trust calibration. Each direction is motivated by a specific limitation or design assumption of the framework that current knowledge does not yet adequately address.

\subsubsection{Review Quality}

A primary research need is to redefine review quality for hybrid human–AI settings. 
A substantial body of prior work on empirically analyzing the usefulness of human review comments~\cite{Bosu2015, Rahman2017} and automated quality evaluation of human comments~\cite{Yang2023, Ahmed2025} provides an important foundation.
Çağlar et al.~\cite{ccauglar2025automated} advance this foundation by introducing a nine-label taxonomy distinguishing six fine-grained review comment smells from three constructive intent categories, and demonstrating through LLM-based classification on 448 labeled comment-diff pairs that intent-boundary labels such as Actionable and Praise are reliably detectable, whereas verification-dependent smells such as Incorrect and Redundant remain near-zero in zero-shot settings without thread-level context.
However, when it comes to AI generated comments studies are limited.
While some studies empirically assess the usefulness and applicability of AI-generated comments~\cite{Cihan2025}, automated evaluation approaches could be further leveraged to improve their quality, but existing automated methods largely focus on evaluating human-written comments.
A key future direction is to develop methods for automatically filtering AI-generated comments based on quality, ensuring that AI-assisted reviews contribute to defect detection, reviewer calibration, and downstream code quality rather than merely increasing comment volume. 
This requires evaluation designs that compare human-only, AI-only, and hybrid review settings, while measuring whether AI support helps reviewers focus on semantically important issues or instead amplifies shallow review behavior~\cite{Tufano2025}.
Critically, any such comparison must be stratified by review-comment type.
Beller et al.~\cite{Beller2014} report that 75\% of review-induced changes are maintainability-related and 25\% address functional defects, and the two categories impose qualitatively different demands on reviewer comprehension.
An evaluation that aggregates across types risks attributing speed gains to AI assistance that in fact reflect a shift in the maintainability-to-functional mix rather than improved defect detection.

In addition, the rise of AI code generation has driven the development of numerous benchmarks, such as SWE-Bench~\cite{jimenez2024swebench}. While these benchmarks are strong indicators of overall code generation performance, their applicability to ad hoc fixes in code review settings remains limited.
Fix generation should be studied as a review-time activity rather than solely as a stand-alone program repair problem. Earlier studies~\cite{marginean2019sapfix,bader2019getafix} have demonstrated the feasibility of automated patch generation, and recent work shows the feasibility of using these systems in actual settings~\cite{frommgen2024resolving,endres2022repair}.
The key question is not just whether a suggested patch compiles, but whether it preserves implementation intent, addresses the reviewer’s concern with minimal collateral changes, and remains easy for authors to audit. 
This highlights the need for benchmarks specifically designed for review-time code generation, where different priorities apply. 
Such systems must be sensitive to comment-conditioned repair, patch minimality, explanation quality, and decision policies that determine when to propose a fix directly versus defer to human revision.

\subsubsection{User Experience for AI-Assisted Review}

We believe the future of code review lies in AI-assisted processes. 
However, the effectiveness of such systems is likely to depend on interface design. 
Prior studies already suggest that reviewers may require different workflows~\cite{Pascarella2018,esem2025rethinking}. 
Accordingly, research should move beyond model accuracy and examine which interaction patterns best support reviewer cognition—for example, inline annotations versus conversational panels, or proactive warnings versus on-demand explanations.
Controlled experiments and field deployments should evaluate not only user preference, but also review latency, mental workload, trust calibration, and whether reviewers actively verify AI outputs before acting on them. 
As with any user-facing system, careful design of the interface is essential. 
While current tools largely rely on chat-based interactions, an effective paradigm for AI, it remains unclear whether this is the optimal user experience for human reviewers.

\subsubsection{Context Enrichment}

Context enrichment is another central research problem, since review quality depends on whether the system can retrieve and organize the right evidence before reasoning begins.
Similar to human review, which is sensitive to a lack of information~\cite{Pascarella2018, partachi2022aide} and fragmented rationale, LLMs are also sensitive to retrieval failures and long-context degradation~\cite{Liu2024}.
Future work should investigate how to combine issue links, historical PR discussions, architectural documentation, ownership data, runtime traces, and PR-issue alignment signals into compact and queryable review contexts.
A key question is whether better performance arises from larger raw context windows, hierarchical retrieval pipelines, stage-specific memory, or explicit intermediate summaries that expose provenance and uncertainty to the reviewer.

In addition, when constructing context, prior code reviews can be treated as a source of information. Code reviews are a rich source of collective project memory~\cite{partachi2022aide}.
However, most current systems do not transform review outcomes into reusable knowledge for later development.
Future research should examine whether post-review summaries can capture accepted rationale, rejected alternatives, risk signals, and reviewer concerns in a form that is both human-readable and machine-retrievable.
If successful, such retrospective memory could support later code generation, reviewer onboarding, consistency checking across related PRs, and the reuse of prior review knowledge without forcing teams to rediscover already settled design decisions.

\subsubsection{Technical Optimization}
PRs vary widely in their intent, scope, and risk, from small documentation fixes to large architectural changes.
Applying a single, uniform automation strategy to all PRs is inefficient, as it can waste resources and distract reviewers with unnecessary analysis. 
For instance, running a complex risk analysis on a simple typo adds little value while increasing computational cost and cognitive load. 
To address this, research should focus on developing clear taxonomies of PRs and classifiers that can distinguish between different types of changes—such as feature additions, bug fixes, and refactoring. This would enable adaptive workflows that apply the appropriate level of automation based on the specific characteristics of each PR.

At the same time, multi-agent review systems introduce the risk of error propagation, where small mistakes in early stages can spread and amplify throughout the pipeline. 
For example, if an issue-linking component incorrectly associates a PR with the wrong issue, this error can mislead later stages such as reviewer recommendation or PR augmentation. 
The result may appear coherent but be disconnected from the actual code changes. 
Future work should focus on designing formal validation methods and consistency checks that allow these components to identify semantic drift and stop the process when needed. Such safeguards are essential to prevent minor errors from evolving into larger, costly failures in automated review systems.

\subsubsection{Evaluation Metrics}
Contemporary software engineering measurement frameworks predominantly prioritize velocity, relying on temporal indicators such as "time-to-merge" or "cycle time" to benchmark performance. 
However, within an AI-augmented workflow where code generation is significantly accelerated, these speed-centric metrics become insufficient and potentially misleading proxies for system efficacy, as they capture phase-level speed rather than end-to-end workflow quality. 
A rapid merge rate is counter-productive if it correlates with a high density of unchecked, autogenerated code that lacks integrity. 
Instead, evaluation should account for how thoroughly the review process is conducted, moving beyond LGTM smells~\cite{Doan2022,Gon2024}.
This shift is particularly important as development increasingly moves toward agent-generated PRs, with code review serving as the primary QA gate where human oversight remains critical. 
Accordingly, future research should focus on designing multi-dimensional evaluation frameworks that move beyond velocity and explicitly capture the depth, rigor, and effectiveness of the review process, even when these dimensions trade off against raw speed.

\section{Conclusion}
\label{sec:conclusion}
Code review has evolved from early practices into modern lightweight PR-based workflows, where it serves as a QA mechanism that extends beyond defect detection to encompass the enforcement of coding standards, developer mentoring, and knowledge sharing. Currently performed through PRs on VCS platforms such as GitHub, GitLab, and Bitbucket, this process integrates collaborative discussion with automated CI/CD pipelines and bot-driven checks. However, the effectiveness of code review is frequently hindered by inherent challenges, such as missing context, high cognitive load, and code review smells. Furthermore, these challenges are exacerbated by recurring bad practices, such as incomplete PR descriptions and poor PR reviewer assignments. Together, these compounding issues ultimately accumulate process debt and diminish the essential benefits of code review. Although recent advancements in LLMs have begun to evolve code review by transforming individual stages, current research and existing tools remain focused on these smaller isolated tasks. By missing the bigger picture, isolated research overlooks how bottlenecks and bad practices are inherently caused by the gaps between individual review stages, necessitating a broader understanding of the entire code review workflow in the era of AI.

To bridge these critical gaps, we propose a visionary AI-powered code review framework that integrates specialized LLM agents with HITL oversight for the entire code review lifecycle. Building upon the foundational insights of prior fragmented studies, we design this framework to connect isolated advancements into a single picture, providing a comprehensive blueprint for a code review workflow that reduces coordination costs while rigorously preserving human authority. The framework begins with the \textit{PR Creation stage}, which automates description generation, establishes issue links, and performs an initial review before PR creation. Following this, the \textit{PR Augmentation stage} performs analysis on the PR, including alignment, risk, runtime, impact, and summary generation, to ground subsequent evaluation. The \textit{Reviewer Selection tool} then suggests PR reviewers utilizing socio-technical workload data. During the \textit{AI-Assisted Code Review stage}, manual inspection is transformed into an interactive task that introduces another layer of abstraction, supported by explanations, automated review requests, fix suggestions, and evaluations of comment usefulness and toxicity. Finally, the \textit{PR Retrospective stage} captures code review process metrics and converts review outcomes into queryable repository memory for the continuous improvement of the workflow.

Looking forward, realizing this AI-powered code review framework opens concrete future directions for software engineering, requiring a structural change in practitioner roles and the development of new metrics for code review. For practitioners, this transition implies a structural change in stakeholder responsibilities, where PR reviewers must evolve from manual inspectors into supervisory operators of agents. Simultaneously, the research community should focus on the creation of new, dynamic evaluation metrics and frameworks specifically tailored to assess multi-agent systems in collaborative settings, moving beyond traditional benchmarks. In particular, implementing these automated systems requires managing underlying socio-technical limitations. Stakeholders should address structural dangers before full deployment, ensuring system transparency, data privacy, and accountability for automated decisions. Furthermore, organizations should actively mitigate the risks of automation bias to prevent the deterioration of team knowledge sharing and implicit mentoring.

To summarize, this study presents a visionary AI-powered code review framework for navigating the future of code review systems amidst the ongoing AI paradigm shift. By fundamentally transforming code review to integrate specialized agents and consolidating fragmented, single-stage advancements and studies into a cohesive workflow, the proposed framework aims to address the persistent friction and bottlenecks characteristic of contemporary code review systems. We hope this paper inspires both practitioners and researchers to look beyond isolated tools and research, and see the bigger picture of software development currently underway.

\bibliographystyle{ieeetr}
\bibliography{references}
\end{document}